\documentclass[
superscriptaddress,twocolumn,
 amsmath,amssymb,
 aps,
prb,
]{revtex4-2}

\usepackage{dsfont}
\usepackage{graphicx}
\usepackage{dcolumn}
\usepackage{bm}
\usepackage{hyperref}
\hypersetup{
colorlinks=true,
	linkcolor=nrppurple,
	citecolor=nrppurple,
	urlcolor=nrppurple,
}
\usepackage{appendix}
\usepackage{fancyhdr}

\newcommand{\ket}[1]{\left| #1 \right>} 
\newcommand{\bra}[1]{\left< #1 \right|} 
\newcommand{\braket}[2]{\left< #1 | #2 \right>} 

\def \Tr{\,\textrm{Tr}\,}

\begin{document}
\definecolor{nrppurple}{RGB}{128,0,128}

	\title{Probing Electromagnetic Nonreciprocity\\ with Quantum Geometry of Photonic States }
\author{Ioannis Petrides}
\affiliation{College of Letters and Science, Physical Sciences Division, University of California, Los Angeles, USA}
\author{Jonathan B. Curtis}
\affiliation{College of Letters and Science, Physical Sciences Division, University of California, Los Angeles, USA}
\author{Marie Wesson}
\author{Amir Yacoby}
\affiliation{Department of Physics, Harvard University,Cambridge,  MA 02138, USA}
\author{Prineha Narang}
\affiliation{College of Letters and Science, Physical Sciences Division, University of California, Los Angeles, USA}

	\date{\today}
	\begin{abstract}  
		Reciprocal and nonreciprocal effects in dielectric and magnetic materials provide crucial information about the microscopic properties of electrons.
		However, experimentally distinguishing the two has proven to be challenging, especially when the associated effects are extremely small.
		To this end, we propose a contact-less detection using a cross-cavity device where a material of interest is placed at its center. 
        We show that the optical properties of the material, such as Kerr and Faraday rotation, or, birefringence, manifest in the coupling between the cavities' electromagnetic modes and in the shift of their resonant frequencies.  
		By calculating the dynamics of a geometrical photonic state, we formulate a measurement protocol based on the quantum metric {and quantum process tomography} that isolates the individual components of the material's complex refractive index and minimizes the quantum mechanical Cram\'er-Rao bound on the variance of the associated parameter estimation.
		Our approach is expected to be applicable across a broad spectrum of experimental platforms including Fock states in optical cavities, or, coherent states in microwave and THz resonators.
	\end{abstract}

\maketitle
Quantum materials offer new technological opportunities while posing key challenges for existing characterization methods.
Phenomena such as superconductivity, topology~\cite{petrides2022semiclassical,zhao2020axion,Curtis.2023a,Narang.2021,Nenno.2020}, magnetism, and collective motion~\cite{Vool.2021,Varnavides.2020} are all manifestations of quantum effects in solid-state systems, which can in turn offer potentially novel electronic device functionalities.  
A commonality between these examples is the way that different symmetries are broken, and the manifestation of these broken symmetries in macroscopic electrodynamic response~\cite{basov2011electrodynamics}. 
One of the most fundamental of such symmetries is time-reversal symmetry (TRS), which when present ensures that material responses are reciprocal, as seen from Onsager's famous relations~\cite{potton2004reciprocity,asadchy2020tutorial}.
The breaking of TRS then allows for nonreciprocal material responses, which are of practical importance for the design of optical and microwave components such as photon routers and circulators~\cite{potton2004reciprocity,asadchy2020tutorial}.  
In topological insulators and semimetals, TRS breaking is expected to induce interesting nonreciprocal responses manifesting as a nonzero Hall conductivity \cite{yu_quantized_2010, da_silva_neto_weyling_2019}, and in correlated insulators is often associated with the onset of magnetic order such as ferromagnetism or antiferromagnetism \cite{corr_kuiri_spontaneous_2022, corr_lee_theory_2019, corr_liu_tunable_2020}.

In recent years there has been an immense interest in unconventional superconductors which spontaneously break TRS~\cite{ghosh2020recent} as they may be candidates for the highly sought after chiral topological superconductivity~\cite{Read.2000,kallin_chiral_2016}.
Signatures of reciprocity breaking can, hence, provide insights to the underlying pairing mechanism, as well as elucidate the coexistence of superconductivity and magnetism~\cite{Poniatowski.2022a,Curtis.2022a}.
Beyond the conventional U(1) gauge symmetry, unconventional superconductors may spontaneously break additional symmetries, such as orbital or spin rotation symmetries~\cite{Poniatowski.2022b,Curtis.2023b}. 
In general, these effects are often orders of magnitude smaller than what can be measured with conventional optical measurements~\cite{xia_high_2006}.
Therefore, estimating the degree by which a material breaks reciprocity requires sensitive apparatuses.

Most frequently, high precision measurements {of non reciprocity} are reported through muon spin relaxation ($\mu$SR) where spin polarized muons precess depending on the complex refractive index and decay in spin-dependent trajectories~\cite{grinenko_superconductivity_2020, mielke_time-reversal_2022, grinenko_split_2021}.
Magneto-optical Kerr probes have also been used to directly demonstrate nonreciprocity at the onset of superconductivity below the critical temperature by measuring the rotation of light polarization with a sophisticated zero-area Sagnac interferometer; in this way reciprocal effects, such as birefringence, are explicitly canceled~\cite{xia_high_2006}. 
These techniques have proven to be very powerful for measuring single crystals~\cite{wei_interplay_2022, schemm_observation_2014} and superconducting/ferromagnetic hybrid materials~\cite{gong_time-reversal_2017}. 
However, the discovery of unconventional superconducting states in van der Waals (vdW) superconductors, such as magic-angle twisted bilayer graphene~\cite{cao_unconventional_2018} and monolayer WTe2~\cite{sajadi_gate-induced_2018, fatemi_electrically_2018}, requires reimagining probes that reach a high level of precision and overcome small sample mode volumes or low densities.

\begin{figure}[t!]
	\centering
\includegraphics[width=1\linewidth]{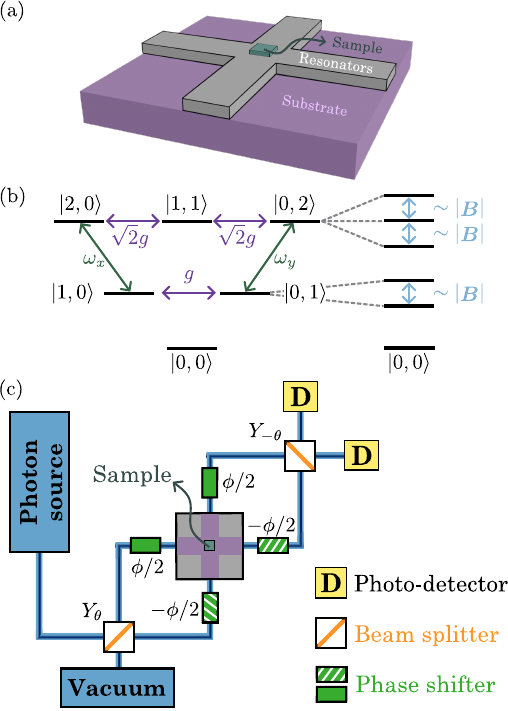}
	\caption{
	(a) A realization of the cross-cavity device using rectangular resonators (gray) fabricated on a substrate (purple).
	The sample (turquoise) is mounted at the intersection of the two resonators and couples evanescently to the electric field modes.
  (b) The spectrum of the two cavities has equispaced energy levels separated by the resonant frequency $\omega_x$ or $\omega_y$.
	The complex coupling $g = g_\chi + i g_\sigma$ between the energy levels is determined by the off-diagonal components of the susceptibility tensor $\chi$ and Hall conductivity $\sigma_H$.
	The energy levels in each Fock subspace are split according to the magnitude of $\bm B$.
	(c) Schematic of the setup. 
	A photon source is used to excite one arm of the cross-cavity device.
	The output electric fields from the two cavities are collected and sent to their corresponding photo-detector.
	In addition, a pre- and post-processing setup of beam splitters and phase shifters is used to rotate the frame of reference to obtain the geometrical state of Eq.~\eqref{eq:CS}.
		}
	\label{fig:fig1}
\end{figure}

Here, we propose an alternative platform to measure the components of the complex refractive index in parallel and provide a detection protocol for disentangling distinct symmetry classifications. 
Specifically, we consider two cross-aligned, single-mode cavities where a sample placed at the intersection is evanescently coupled to the electromagnetic fields.
We calculate the evolution of photonic states and relate the photon occupation number to the quantum metric characterizing the space of states.
As the latter is determined by the sample's susceptibility and conductivity, the induced quantum geometry is used to separate reciprocal and nonreciprical effects, in addition to minimizing the quantum uncertainty of the measured parameters.
Finally, we define an optimised detection protocol that uses the minimum number of sampling points to extract the sample's optical properties.
Notably, our contact-free spectroscopic probe is particularly useful for studying materials where obtaining reliable electrical contacts can be challenging, such as in the aforementioned vdW 2D materials. 
{Furthermore, our proposal can be generalized for both coherent and Fock states, allowing for various implementations across the optical, terahertz, and microwave regimes.}

\section{Results}
\subsection{Cross-cavity model}
We consider a cross-cavity device in a planar geometry with a small dielectric sample at the intersection, see Fig.~\ref{fig:fig1}(a).
The latter is well described by a $\delta(\bm r)$ distribution, susceptibility tensor $\chi_{ij} $, and a conductivity tensor $\sigma_{ij} = \varepsilon^{ij}\sigma_H$, where $\sigma_H$ is the Hall conductivity and $\varepsilon^{ij}$ is the Levi-Civita tensor. 
{For simplicity, the diagonal conductivity is set to zero and its effect is incorporated in the coherence time of the device.}
The cavities are characterized by a conductivity tensor $\sigma_0 $ and a susceptibility tensor $\chi_0$ which define the ``reference vacuum" for the electric field.
Without loss of generality, we choose a trivial reference vacuum with zero conductivity $\sigma_0 = 0$ and an isotropic susceptibility tensor $\chi_0\sim\chi_0\mathds{1}$; additional contributions from nontrivial vacua can be equally treated by absorbing them into the definitions of $\delta(r)$, $\chi$ and $\sigma$.
Furthermore, we assume that each cavity can support a single mode with the electric field sufficiently permeating into free space such that it evanescently couples to the sample.

Before introducing quantum mechanical effects, it is instructive to showcase the behaviour of the device in its classical limit.
Solutions to Maxwell's equations can be obtained perturbatively using classical electromagnetic fields with their evolution computed using a standard Green's function approach (see Methods \ref{apx:Classical}).
As a result, the action of the sample becomes equivalent to a beam splitter where an incident electromagnetic field scatters to the available channels;
{in this geometry, the associated split ratio is determined by the magnitude of the off-diagonal component of the
complex refractive index, while a relative phase shift between the two arms of the device will only occur when there is a finite imaginary off-diagonal element.
}
Importantly, the classical treatment of Methods \ref{apx:Classical} is only perturbatively valid with leading-order corrections proportional to $\sigma_H / \omega\ll 1$ {(note that in natural units conductivity and frequency both have units of energy, see Methods \ref{apx:Classical})}.
Hence, detecting nonreciprocal effects may be challenging due to background radiation or experimental uncertainties.

We now treat the system quantum mechanically {in the case of closed dynamics, i.e., in the absence of any coupling to the environment; we will later introduce nonunitary process, e.g., losses, by replacing the unitary evolution operator with a completely positive map.}
The relevant quantised Hamiltonian in the rotating wave approximation is given by (see Methods~\ref{appx:Hamiltonian})
\begin{eqnarray}
	\hat{H} = \Delta \tilde{\omega} (\hat{a}^\dagger_y \hat{a}_y - \hat{a}_x^\dagger \hat{a}_x ) + g \hat{a}_x^\dagger \hat{a}_y + \textrm{h.c.}
	\label{eq:RWA-boson}
\end{eqnarray}
where $2\Delta \tilde{\omega} = \tilde{\omega}_x - \tilde{\omega}_y $ is the difference of the cavities' resonant frequencies, with $\tilde{\omega}_i = \omega_i+\delta\omega_i$ the sum of the bare resonant frequency of the cavity in the $i$th direction $\omega_i$ (assumed to be almost equal in the two directions) and the shift ${\delta\omega_i}$ due to the diagonal terms of the sample's susceptibility $\chi^{xx}$, or $\chi^{yy}$.
The hybridization between the two cavity modes $g = g_\chi + ig_\sigma$ is determined by the real coupling $g_\chi$ induced by a finite off-diagonal susceptibility $\chi^{xy}$, and the imaginary coupling $g_\sigma$ induced by a finite Hall conductivity $\sigma_H$.
Hamiltonian \eqref{eq:RWA-boson} can be interpreted as the effective dynamics of a spin in a magnetic field $\bm B$, namely
\begin{eqnarray}
	\hat{H} =\bm B\cdot \hat{\bm S}
	\label{eq:RWA}
\end{eqnarray}
where $ \hat{\bm S} = \{\hat{S}_x,\hat{S}_y, \hat{S}_z\}$ define the elements of the SU(2) algebra and are given in terms of the creation and annihilation operators $\hat{a}_i$ and $\hat{a}^\dagger_i$ (see Methods~\ref{appx:Hamiltonian}), and
\begin{eqnarray}
    \begin{array}{cc}
      \bm B = \{g_\chi, g_\sigma, \Delta\tilde{\omega}\}.
    \end{array}
\end{eqnarray}
is a vector determined by the complex coupling $g$ between the two cavities and their relative frequency difference $\Delta \tilde{\omega}$.
While both the $\bm B_x$ and $\bm B_z$ components are expected due to the polarizability of the sample, mode splitting, or even due to geometrical effects of the device's shape and impurities, a nonzero $\bm B_y$ component arises only when time-reversal symmetry is broken.

Since the Hamiltonian \eqref{eq:RWA} conserves the total photon number operator $\hat{N} = \hat{ a}_x^\dagger\hat{ a}^{\empty}_x+\hat{ a}^\dagger _y\hat{ a}^{\empty}_y$, the Fock space is diagonal with respect to the total number of photons  $N= \langle \hat{N}\rangle$ in the cross-cavity system.
In the absence of any coupling, i.e., when $ |\bm B|= 0$, the spectrum of the system is given by the tensor product of equispaced energy levels corresponding to each cavity, see Fig~\ref{fig:fig1}(b). 
A finite coupling between the modes or a non-zero energy difference between the cavities' resonant frequencies lifts the degeneracy in each Fock subspace and forms an effective spin-$N/2$ Schwinger boson in a magnetic field $\bm B$ with the irreducible representations of the $SU(2)$ algebra characterized by the total number of photons~\cite{mathur2010n}.

\begin{figure}[t!]
	\centering
	\includegraphics[width=1\linewidth]{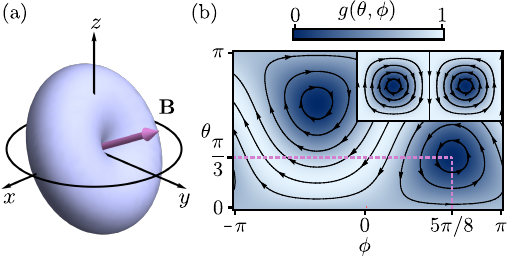}
	\caption{
	 (a) Polar plot of the quantum metric $g(\theta,\phi)$ as a function of $\theta$ and $\phi$.
  The quantum metric vanishes along the vector $\bm B$, chosen randomly at $(\theta_B,\phi_B)=(\pi/3,5\pi/8)$.
  (b) Density plot of the quantum metric and the precession trajectories of the state vector $\bm S$.
  The inset shows the trajectories when $\bm B$ is along the $y$ axis.
		}
	\label{fig:fig2}
\end{figure}

\subsection{Unitary evolution}
Our measuring protocol is based on observing the dynamics of the $N$-photon geometrical Fock state
\begin{eqnarray}
\begin{array}{c}
 \ket{\psi_0(\theta,\phi)} = \frac{1}{\sqrt{N!}}\left(\cos(\frac{\theta}{2})e^{-i\frac{\phi}{2}}\hat{a}_x^\dagger +\sin({\frac{\theta}{2}})e^{i\frac{\phi}{2}}\hat{a}_y^\dagger  \right)^N\ket{0,0}
\end{array}
\label{eq:CS}
\end{eqnarray}
where $\ket{0,0}$ is the vacuum state with zero photons in both cavities, and $\ket{n,m} =\frac{1}{\sqrt{n!m!}}( \hat{a}^\dagger _x)^n (\hat{a}^\dagger _y)^m\ket{0,0}$ represents the $(n+m)$-photon eigenstate with $n$ ($m$) photons in the $x$ ($y$) cavity.
The proposed geometrical state $\ket{\psi_0(\theta,\phi)}$ can be prepared by a pre-processing optical setup of beam splitters and phase shifters, see~Fig.\ref{fig:fig1}(c). 
Specifically, a photon source initializes the system in the $\ket{N,0}$ Fock state of $N$ photons in the {$x$ cavity}.
The beam splitter rotates the state according to the operator $Y_\theta = e^{i{\theta}\hat{S}_y}$, with $\theta$ defined by the split ratio.
Finally, a phase shifter is used to further rotate the state by $Z_\phi  = e^{i{\phi}\hat{S}_z} $, leading to the desired geometrical state $\ket{\psi_0(\theta,\phi)} = Z_\phi Y_\theta\ket{N,0}$ of Eq.~\eqref{eq:CS}.

The geometrical photonic state in the device will evolve according to
\begin{eqnarray}
    \ket{\psi_t(\theta,\phi)} = U(t)\ket{\psi_0(\theta,\phi)} 
\end{eqnarray}
where $U(t) = e^{i t \hat{H}}$ is the evolution operator;
consequently, the state undergoes precession around the vector $\bm B$ with frequency proportional to $|\bm B|$.
However, the direction and magnitude of $\bm B$ are a priori unknowns, therefore, the precision in estimating the angular change is limited by the quantum metric $g(\theta,\phi)$ which determines the distance between adjacent quantum states, namely ${|\braket{ \psi_{t +\delta t}(\theta,\phi)}{ \psi_t(\theta,\phi)}| = 1 -\frac{1}{2}\delta t^2 g(\theta,\phi)}$, and serves as a measure of their distinguishability~\cite{cramer1999mathematical,rao1992information,sidhu2020geometric,yu2022quantum}.
For our system, the quantum metric is given by
\begin{eqnarray}
    g(\theta,\phi) =1-(\underline{\bm B}\cdot \underline{\bm S})^2
\end{eqnarray}
where $g(\theta,\phi)\in[0,1]$ takes values between zero and one [see Fig.~\ref{fig:fig2}(a)], ${\underline{{\bm B}} = \frac{\bm B}{|\bm B|}}$ is the normalized vector and
\begin{eqnarray}
\underline{\bm S} = \frac{1}{N}    \langle \hat{\bm S}\rangle = \{\sin(\theta)\cos(\phi), \sin(\theta)\sin(\phi),\cos(\theta)\}
\end{eqnarray}
is a unit vector on the 2-sphere defined by the expectation value of the spin operators $\hat{\bm S}$ with respect to the geometrical state of Eq.~\eqref{eq:CS}.
For example, when $\underline{\bm S}$ is perpendicular to $\underline{\bm B}$, the quantum metric is maximized and the state undergoes precession around the equator defined by $\underline{\bm B}$, see Fig.~\ref{fig:fig2}(b).
On the contrary, when the initial state vector $\underline{\bm S}$ is parallel to $\underline{\bm B}$ the quantum metric is equal to zero and the state will not undergo precession.

\begin{figure*}[ht]
	\centering
	\includegraphics[width=\textwidth]{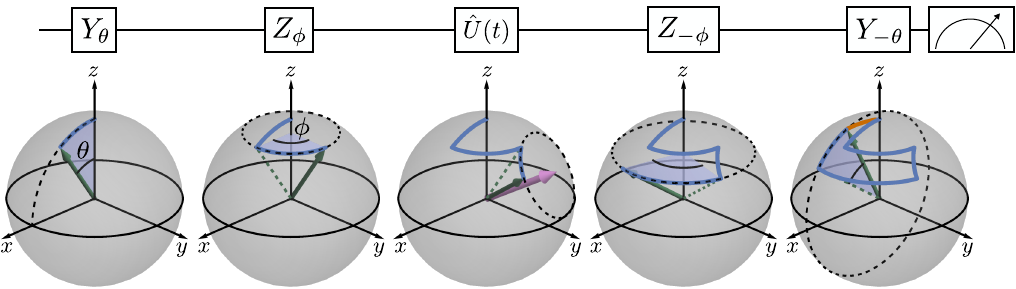}
	\caption{	The measuring sequence. The initial state is represented by a vector pointing at the north pole of the Bloch sphere.
 The first beam splitter induces a rotation around the y-axis by an angle $\theta$.
 The phase shifters further rotate the state vector around the z-axis by an angle $\phi$.
 The state precesses around $\bm B$ (pink vector) according to the evolution operator $\hat{U}(t)$.
 Next, phase shifters are used to rotate the state around the z-axis by an angle $-\phi$.
 The last beam splitter induces a rotation around the y-axis by an angle $-\theta$.
 Finally, the state is projectively measured by the photo-detector.
 The change between the initial and final state vectors is determined by the quantum metric, indicated here by the orange line.
		}
	\label{fig:fig3}
\end{figure*}

\begin{figure}[ht]
	\centering
	\includegraphics[width=1\linewidth]{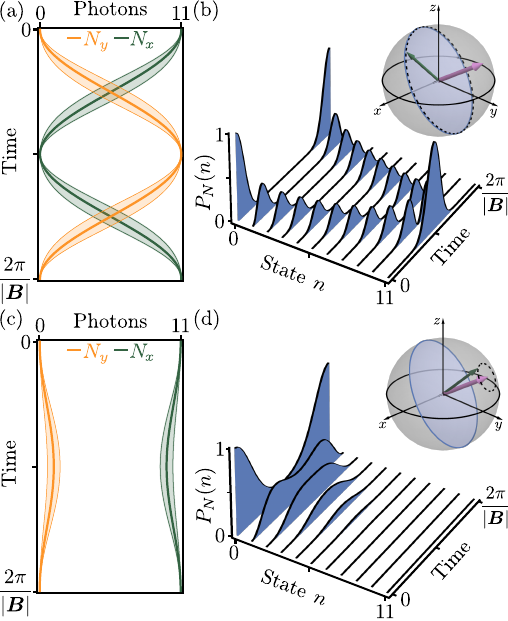}
	\caption{ Evolution of the $N=11$ excitation manifold of states for two different values of $\theta$ and $\phi$.
	 (a) The average photon number in each cavity as a function of time for $\bm S$ almost perpendicular to $\bm B$, i.e., $g(\theta,\phi) \approx 1$.
    The width of each line indicates the variance $\Delta N_i ^2$.    
  The average photon number is transfered entirely between the two cavities with a vanishing variance at period $T = 2\pi/ |\bm B|$ and $T/2$.
  (b) The probability distribution $P_{N}(n)$ of each state associated to (a).
  The photon population oscillates between the entire set of states with frequency $ |\bm B|$. 
  In addition, the precession path (dashed line) associated to the probability evolution is shown on the Bloch sphere. 
  The blue region is a plane perpendicular to $\bm B$.
  	 (c) The average photon number in each cavity as a function of time for $\bm S$ almost parallel to $\bm B$, i.e., $g(\theta,\phi) = 0.1$.
    The width of each line indicates the variance $\Delta N_i ^2$.    
  In this case, the average photon number stays primarily in the $x$ cavity throughout the cycle.
    (d) The probability distribution $P_{N}(n)$ associated to (c). 
  The photonic states oscillate between only a certain subset with frequency $|\bm B|$. 
  The precession path (dashed line) on the Bloch sphere associated to the probability evolution is now located closer to the position of the vector $\bm B$.
		}
	\label{fig:fig4}
\end{figure}

We extract the quantum metric by performing a projective measurement of the final state after a post-processing setup of a beam splitter $Y_{-\theta} $ and phase shifters $Z_{-\phi} $, cf.~Fig.~\ref{fig:fig1}(c).
The measuring protocol, shown in Fig.~\ref{fig:fig3}, is based on the algebraic properties of the $SU(2)$ group which, ultimately, are related to the geometry of the photonic states.
The probability distribution of simultaneously measuring $N-n$ photons in the {$x$ cavity} and $n$ photons in the {$y$ cavity} after evolving the state for time $t$ is given by (see Methods~\ref{appx:algebra} for specific expression)
\begin{eqnarray}
   {P_{N}{(n)} =| \bra{N-n,n}  U' (t)\ket{N,0} |^2}\,,
\end{eqnarray}
where $U'(t) = Y_{-\theta} Z_{-\phi}U(t)Y_{\theta} Z_{\phi}$ is the evolution operator in the rotated basis induced by the pre- and post-processing optical setup.
The mean photon number in each cavity $N_i = \langle \hat{a}^\dagger_i\hat{a}_i\rangle$ is given by 
\begin{eqnarray} 
N_y =N-N_x= N g(\theta,\phi)\sin^2 \left(\frac{ |\bm B|t}{2} \right)
    \label{eq:PDe}
\end{eqnarray}
 and the variance by ${\Delta N_i ^2 =N_y - \frac{1}{N}N_y ^2}$ for both $i=x$ and $y$.
Depending on the quantum metric, the average occupation of each cavity will oscillate with frequency $ |\bm B|$; these oscillations are maximized (minimized) when the initial state vector $\underline{\bm S}$ is perpendicular (parallel) to $\underline{\bm B}$. 
The direction and magnitude of $\bm B$ can, therefore, be determined by observing the precession of the geometrical state for different values of $\theta$ and $\phi$.
When the initial state vector $\bm S$ is prepared to be perpendicular to $\bm B$, the Fock states $\ket{N-n,n}$ of the $N$-excitation submanifold will perform oscillations with frequency $ |\bm B|$ and the entire photon population will be transfered between the two cavities, cf. Fig.~\ref{fig:fig4}(a) and (b).
Notably, the variance of the mean photon number becomes zero every half oscillation period. 
On the other hand, when the state vector $\underline{\bm S}$ lays almost parallel to $\bm B$, the photons remain primarily in the $x$ cavity, cf. Fig.~\ref{fig:fig4}(c) and (d).

\subsection{Process tomography of nonunitary evolution}
In any experimental setup, the system will decohere through various decay channels due to its coupling to the environment {and due to Ohmic dissipation (finite conductivity).
We, hence, replace }the unitary evolution by a completely positive map $\mathcal{E}_t$ that evolves the initial density matrix ${\rho_0(\phi,\theta) = \frac{1}{{N!}}( Z_\phi Y_\theta \hat{a}^\dagger _x\hat{a} _x   Y_{-\theta} Z_{-\phi})^N}$ of a Fock state according to $\rho_t (\phi,\theta)  = \mathcal{E}_t [\rho_0(\phi,\theta)  ]$.
For process tomography of a noisy implementation of our unitary gate $U(t)$ there is a total of sixteen free parameters that have to be uniquely determined~\cite{takahashi2013tomography,chuang1997prescription}.
Here, we assume that the dominant contributions to $\mathcal{E}_t $ are well captured by a photon leakage out of the device with coherence time $\tau$ that results in a nonunitary evolution towards the center of the Bloch sphere.
Such process has four unknowns that can be extracted using three states and a set of positive operator valued measure (POVM) that consists of two elements $\{\hat{N}_x, \hat{N}_y\}$.
The states are chosen as $\rho^{(1)} = \rho_t(0,0)$, $\rho^{(2)} = \rho_t(\pi/2,0)$, and $\rho^{(3)} = \rho_t(\pi/2,\pi/2)$ with associated photon numbers $\{N_x ^{(i)} , N_y ^{(i)} \}$.
The three components of $\bm B$, and the coherence time $\tau$ can be found from a minimization routine of the relations
\begin{eqnarray}
\begin{array}{rl}
e^{-t/\tau}&={\frac{1}{3N}{\sum\limits_i (N_y ^{(i)}+ N_x ^{(i)})}}\\
\sin^2\left(\frac{|\bm B| t}{2}\right) &={\frac{1}{2N}{\sum\limits_i N_y ^{(i)}}}\\
\tan\theta_{\bm B}& =  \sqrt{\frac{2N_y ^{(1)}}{-N_y ^{(1)}+N_y ^{(2)}+N_y ^{(3)}}}\\
\tan\phi_{\bm B} &=\sqrt{\frac{N_y ^{(1)}+N_y ^{(2)}-N_y ^{(3)}}{N_y ^{(1)}-N_y ^{(2)}+N_y ^{(3)}}}
\end{array}
\end{eqnarray}
where $\theta_{\bm B}$ ($\phi_{\bm B}$) is the polar (azimuthial) angle of $\bm B$.

\section{Discussion and outlook}

We highlight that our protocol can be implemented in optical Fabry-Perot cavities, allowing access to complex dielectric properties at THz and optical frequencies \cite{sturges_quantum_2021, brown_deterministic_2003, velez_preparation_2019}.  
We note that experimentally, components of our proposal have been realized, where microwave cavity devices drive polarization selective transitions~\cite{henderson_high-frequency_2008, alegre_polarization-selective_2007}.
Here, we propose an implementation in the microwave regime, aiming to maximize the sensitivity of this technique. 
Our approach involves utilizing high-quality factor superconducting resonators, either in a coplanar waveguide geometry or in 3D cavities, which can achieve large Q factors ranging from 10$^7$ to 10$^{12}$~\cite{romanenko_three-dimensional_2020}. 
The initial Fock state can be prepared using a coupler transmon that dispersively couples to the two cavity modes, enabling quantum state transfer between the qubit and the cavity~\cite{hofheinz_generation_2008}. 
The coupler transmon functions as a beamsplitter that uses the nonlinearity of the Josephson junction to drive parametric conversion, as described previously \cite{wang_efficient_2020, gao_programmable_2018}. 
Furthermore, by controlling the phase of the microwave drive tones applied to the transmon, the phase between the photonic states can also be manipulated.
In addition, the possibility of using highly entangled states as optimal probes provides a promising route to extract the complex dielectric properties of the material by attaining the Heisenberg limit of precision~\cite{matthews2016towards}.

For such a microwave device at finite temperature, the uncertainty associated to the POVM is bounded by thermal noise.
For a thermal coherent state, the experimental error in measuring the $\bm B_y$ component from the photon expectation value $N_y$ is given by the variance (see Methods~\ref{appx:precesion})~\cite{xia_detection_2015}
\begin{eqnarray}
\begin{array}{rl}
    \Delta \bm B_y ^2& \gtrsim \frac{ \Delta^2 N_y }{\left(\partial N_y / \partial \bm B_y\right)^2}\\
     &  \gtrsim  \frac{2 n_{th.} +h_2 }{\tilde{ N}F}
\end{array}
\end{eqnarray}
where $F=\frac{1}{h_2}(\frac{\partial h_2}{\partial \bm B_y})^2$ is the Fisher information, $\tilde{N}$ is the mean number of photons in the coherent state and $ n_{th.}\approx1/(e^{\beta \omega_0}-1)$ is the mean number of thermal photons, with $\beta$ is the inverse of temperature multiplied by the Boltzmann factor.
In reality, the period of precession ${T=2\pi | \bm B|^{-1}}$ will be dominated by $\Delta \tilde{\omega}$ due experimental challenges in engineering identical cavities; typical values that can be achieved in superconducting cavities operating at GHz frequencies can be as small {as~$\sim1$~MHz}, leading to a precession period on the order of a few $\mu s$, which is well below the device's lifetime $\tau\sim 1s$.  
Assuming a coherent state with a typical mean number of photons $\tilde{N} = 10$ and a thermal photon number $ n_{th.} = 1.6 $ at $ 100$mK, we estimate the optimal mean-squared error (see Methods~\ref{appx:precesion}) to be of order ${\sqrt{\tau}\Delta \bm{B}_y \approx300}\text{Hz}/\sqrt{\text{Hz}}$ or in terms of unitless Hall conductivity $\sqrt{\tau}\Delta  \sigma_H \approx \eta^{-1} 10^{-7} (e^2/h)/\sqrt{\text{Hz}}$, where $\eta$ is a geometrical factor proportional to the ratio of the volume of the sample over the cavity mode volume (see Methods~\ref{appx:Hamiltonian}).
In materials that exhibit notably small Kerr signals, such as Sr$_2$RuO$_4$, low frequency Hall conductivity is reported as $~10^{-2} (e^2/h)$~\cite{lutchyn_frequency_2009}, which is well within our sensitivity. 

To conclude, we discuss possible extensions of the proposal. 
One of the crucial ingredients of our protocol is the coupling of the sample to the evanescent modes of the electric field in the cross-cavity device.
However, such coupling can be hindered by contact imperfections that can arbitrary change the complex susceptibility or generate stray fields.
Therefore, introducing an insulating layer between the sample and the cavities can prevent any build-up of surface effects.
Moreover, Eq.~\eqref{eq:RWA} is valid only in the limit where the ratio $|\bm B|/\omega$ is treated perturbatively, i.e., the cavities' frequencies are much larger than the coupling strength. 
Even though for small values of $|\bm B|$ this is automatically satisfied, we note that the rotating wave approximation will break down in cases where the Hall conductivity $\sigma_H$ or polarizability $\chi$ are large compared to the cavities' central frequency.
In this regime, the system is expected to undergo a phase transition where the ground state acquires a nonzero photon number.
Our technique can also be expanded on to study nonlinear media, which can connect topology with spectral electronic properties~\cite{hayashi_magneto-optical_2021}.

In the proposed platform, we demonstrate how oscillations between cavity modes in a cross-aligned geometry can be used to detect the relative complex dielectric function of a sample placed at the intersection.
The Hamiltonian dynamics describing the photonic states in the cavities is derived, where the sample's Hall conductivity $\sigma_H$ and susceptibility tensor $\chi$ are shown to induce a complex coupling between the two cavity modes, as well as shift their resonant frequencies.
By considering the $N$-photon excitation subspace, we determine the evolution of a geometrical quantum state prepared by a pre- and post-optical setup of beam splitters and phase shifters.
We show that the oscillations of the photon population in each cavity depend on the quantum metric which is defined by both the shift of the resonant frequencies induced by the diagonal terms of the susceptibility $\chi_{xx}$ and $\chi_{yy}$, as well as the complex coupling between the modes induced by the off-diagonal elements of the susceptibility $\chi_{xy}$ and Hall conductivity $\sigma_H$.
Finally, we present a measuring protocol to uniquely determine the dielectric properties of the sample using a minimal number of sampling points.

\section{Methods}

\subsection{Classical treatment}
\label{apx:Classical}

The equation of motion for the vector potential in the cross-cavity geometry up to linear order in the complex susceptibility is given by the Helmholtz equation
	\begin{eqnarray}
	-n^2 \ddot{\bm A} + \nabla^2 \bm A = \delta(\bm r) \mu_0 \sigma  \dot{\bm A}\,,
	\label{eq:Helm}
	\end{eqnarray}
where $n^2 = n^2_0 + \delta n^2 $ is determined by the refractive index of the cavity ${n^2_0 = \mu_0\epsilon_0(1+\chi_0 )}$ and the sample ${\delta n^2  =  \delta(\bm r)\mu_0\epsilon_0 \chi}$.
{Without lost of generality, we take $\epsilon_0 = \mu_0 = 1$ for simplicity and work in natural units where conductivity has units of energy and susceptibility is dimensionless.}
The incident vector potential in each waveguide is classically described by ${\bm A ^{\text{inc.}}_i = \alpha_i e^{i\omega_i t} f_i(\bm r)}$ where $\alpha_i$ is a complex coefficient, $\omega_i$ is the frequency and $f_i(\bm r)$ is the mode profile.
From the equation of motion \eqref{eq:Helm} the function $f_i(\bm r)$ satisfies (where the index $i$ is omitted for simplicity)
\begin{eqnarray}
  {\int d\bm r f^*\nabla^2 f = \int d^3\bm r f \nabla^2 f^*=-(n_0 \omega)^2}
  \label{eq:ModeProfRel}
\end{eqnarray}
as well as $\int d\bm r(|n_0\omega f|^2 + f^* \nabla^2 f) = 0$. 
The total vector potential can be found by decomposing the solutions into incident and scattered fields, i.e., ${\bm A (\bm r) =  \bm {A}^{\text{inc.}}(\bm r) + \bm{A}^{\text{sca.}}(\bm r) }$, and solving Eq.~\eqref{eq:Helm} using a Green's function approach. 
In the regime where $\chi$ and $\sigma$ can be treated perturbatively, the total vector potential at the output of the device is given by 
\begin{equation}
    \bm A (\bm r) = \tau(\bm r, \bm r') \bm A^{\text{inc.}}(\bm r')
\end{equation}
where $\tau(\bm r, \bm r') = G(\bm r, \bm r')(\mathds{1}-Z)^{-1}$ is the transfer matrix, $Z =\int d\bm r\delta(\bm r) (\chi\bar{\omega}^2  + i \sigma\bar{\omega}) $ is related to the complex susceptibility, and $G(\bm r, \bm r')$ is the Green's function of the homogeneous equations of motion
\begin{eqnarray}
    \left((n_0 \bar{\omega})^2 +\nabla^2\right)G(\bm r, \bm r') = \delta(\bm r, \bm r')
\end{eqnarray}
with $\bar{\omega}_{ij} = \omega_i \delta_{ij}$.

\subsection{Quantum treatment}\label{appx:Hamiltonian}

The Lagrangian density corresponding to the differential equations~\eqref{eq:Helm} is given by 
\begin{eqnarray}
	\mathcal{L} = \frac{1}{2}\left(\left(n\dot{\bm A}\right)^2 - \left(\nabla \times\bm A \right)^2 -\delta(\bm r) \bm A\sigma \dot{\bm A}\right)\,,
\end{eqnarray}
with the corresponding Hamiltonian given by
\begin{eqnarray}
	\begin{array}{rl}
	H &= \int d\bm r \left(\bm \Pi \cdot \dot{\bm A} - \mathcal{L}\right)\\
		& =\frac{1}{2} \int d\bm r \left(\left(\bm{\Pi} - \delta({\bm r}) \sigma \bm{A} \right)^T \cdot n^{-2}\cdot \left(\bm{\Pi} - \delta({\bm r}) \sigma \bm{A} \right)\right. \\&\left.+  \left(\nabla \times\bm A \right)^2\right) 
	\end{array}
	\label{eq:Ham}
\end{eqnarray}
where $\bm \Pi =\frac{ \partial \mathcal{L}}{\partial \dot{\bm A}} $ and in the second line we have neglected surface terms which vanish in the limit of large volume.
It is understood that in principle both $n^2$ and $\sigma$ are tensors characterizing the susceptibility and Hall conductivity, respectively, and the transpose acts on the vectorial indices, which in particular will change sign $(\sigma \bm A)^T = -\bm{A}^T \sigma$ due to the antisymmetric nature of the Hall conductivity.

We quantize the Hamiltonian \eqref{eq:Ham} by defining the vector potential as an operator (taking $\hbar =1$)
\begin{eqnarray}
	 A_i \rightarrow \hat{A}_i= \sqrt{\frac{1}{2n^2 _0 \omega_i}}  f_i(\bm r) \hat{ a}_i + h.c.\,,
\end{eqnarray}
where $\omega_i$ and $f_i (\bm r)$ are the frequency and spatial profile of the cavity mode in the $i$th direction, respectively.
Using the relations \eqref{eq:ModeProfRel} and the normalization condition $\int d^3\bm r |f|^2 =1$, the Hamiltonian operator in the rotating-wave-approximation is given by
\begin{eqnarray}
	\hat{H} =\omega_0 \hat{N} + {\Delta \tilde{\omega}}\left(\hat{ a}_y^\dagger\hat{ a}^{\empty}_y-\hat{ a}^\dagger  _x\hat{ a}_x^{\empty} \right) + g   \hat{a}^\dagger _x \hat{a} _y + \text{h.c.}\notag\\
 \label{eq:GRWA}
\end{eqnarray}
where $\hat{N} =\hat{ a}_x^\dagger\hat{ a}^{\empty}_x+\hat{ a}_y^\dagger\hat{ a}^{\empty} _y  $ is the total photon number operator, ${\omega_0 \gg \Delta \tilde{\omega}, |g|}$ are the central frequency, mode splitting and hybridizations, respectively, with ${g = g_\chi + ig_\sigma}$ a complex coupling between the two modes.
The central frequency ${2 \omega_0 = \tilde{\omega}_x + \tilde{\omega}_y}$ and splitting ${2\Delta \tilde{\omega} = \tilde{\omega}_x - \tilde{\omega}_y }$ are determined by the resonant frequencies of the cavities ${\tilde{\omega}_i = \omega_i+\delta\omega_i}$, with $\omega_i$ the bare frequency and ${2\delta\omega_i=\eta\omega_i {\chi^{ii}}}$ the frequency shift due to the diagonal susceptibility $\chi^{ii}$ of the sample. 
The real ${g_\chi=\eta \sqrt{\omega_x\omega_y}{\chi^{xy}}}$ and imaginary $g_\sigma =\frac{ \eta}{2}\frac{\omega_x + \omega_y}{\sqrt{\omega_x\omega_y}}{\sigma_H}$ coupling between the modes arise due to an off-diagonal susceptibility $\chi^{xy}$ and finite Hall conductivity $\sigma_H$, respectively. 
The geometrical factor ${\eta = n_0^{-2}\int d\bm r \delta(\bm r) |f(\bm r)|^2}$ in the above expressions is determined by the cavities' refractive index, mode profile, and sample shape.
When the cavities' bare frequencies are equal, i.e., $\omega_x=\omega_y =\omega_0$, the mode splitting and hybridization are simplified to $ 2 \Delta \tilde{\omega} =\eta\omega_0 (\chi^{yy}-\chi^{xx} )$, and ${g = \eta(\omega_0 \chi^{xy} + i \sigma_{H})}$, respectively.

Equation \eqref{eq:GRWA} describes two bosonic oscillators with commutation relations $\left[\hat{a}_i,\hat{a}^\dagger_j\right] = \delta_{ij}$ and resonant frequencies $\tilde{\omega}_x$ and $\tilde{\omega}_y$, that interact via a complex coupling $g$.
It can be readily recast into Eq.~\eqref{eq:RWA} by defining the elements of the $SU(2)$ algebra as
\begin{eqnarray}
\begin{array}{c}
    \hat{ S}_x =\frac{1}{2}\left(\hat{ a}_x^\dagger\hat{ a}^{\empty}_y+h.c.\right)\,,      \hat{ S}_y =\frac{i}{2}\left(\hat{ a}_x^\dagger\hat{ a}^{\empty}_y- h.c.\right)\,,\\ \text{and}\,
      \hat{ S}_z =\frac{1}{2}\left(\hat{ a}_x^\dagger\hat{ a}^{\empty}_x-\hat{ a}^\dagger _y\hat{ a}^{\empty}_y \right)
\end{array}
\end{eqnarray} 
with commutation relations $\left[\hat S_i , \hat S_j\right] = i \epsilon^{ijk}\hat S_k$ and ${\hat {\bm S}^2 = \frac{1}{4}(\hat N^2 - 2\hat N)}$. 

\subsection{Probability evolution}\label{appx:algebra}

The evolution of the geometrical ground state, c.f., Eq.~\eqref{eq:CS}, is obtained from the operator relations
\begin{eqnarray}
\begin{array}{rl}  
    U(t) a^\dagger _x U^{-1}(t) =&  a^\dagger _x\left( \cos\left(\frac{ |\bm B|t}{2}\right) + i  \sin\left(\frac{ |\bm B|t}{2}\right) \underline{\bm B}_z\right)\\\\&+i a^\dagger _y\sin\left(\frac{ |\bm B|t}{2}\right)\left(   \underline{\bm B}_x - i   \underline{\bm B}_y\right)
\end{array}\\
\begin{array}{rl}
  
    U(t) a^\dagger _y U^{-1}(t) =&  a^\dagger _y\left( \cos\left(\frac{ |\bm B|t}{2}\right) - i  \sin\left(\frac{ |\bm B|t}{2}\right) \underline{\bm B}_z\right)\\\\&+i a^\dagger _x\sin\left(\frac{ |\bm B|t}{2}\right)\left(   \underline{\bm B}_x + i   \underline{\bm B}_y\right)
\end{array}
\end{eqnarray}
The probability distribution of photon states after evolving with $\hat U(t)$ is given by
\begin{eqnarray}\begin{array}{rl}
   P_{N}{(n,n_0)}& =| \bra{N-n,n}  U' (t)\ket{N-n_0,n_0} |^2\\
   & = \sum\limits_{m = 0}^{N-n_0} \sum\limits_{m'=0}^{n_0} C_m^{N-n_0} C_{m'}^{n_0}\left(h_1\right) ^{N-m-m'} \\ &\hspace{70pt}\times\left(h_2\right) ^{m+m'}\delta_{m-n,m'-n_0}
\end{array}
\label{eq:ProbD}
\end{eqnarray}
where $C_m ^n$ is the binomial coefficient and
\begin{eqnarray}
    h_1 = \cos^2\left(\frac{ |\bm B|}{2} t\right) +\big(1-g(\theta,\phi)\big)\sin^2\left(\frac{ |\bm B|}{2} t\right)\\
    h_2 = g\left(\theta,\phi\right)\sin^2 \left(\frac{ |\bm B|}{2} t\right) 
    \label{eq:prob amp}
\end{eqnarray}
In the specific case of $n_0 = 0$, i.e., when the initial state has all photons in the $x$ cavity, equation \eqref{eq:ProbD} is reduced to
\begin{eqnarray}
\begin{array}{c}
     P_{N}{(n)} =C^N _n  \left(h_1\right)^{N-n}\left(h_2\right)^n\,.
\end{array}
\end{eqnarray}

\subsection{Thermal noise }\label{appx:precesion}
We assume a microwave resonator device with intrinsic loss rate $\kappa_i$, driven by a thermal coherent state with external coupling rate $\kappa_{ext.}$. 
In the density matrix representation the state operator of a thermal coherent state is given by
\begin{eqnarray}
    \rho =\sum\limits_{n_x,n_y} \frac{1}{\mathcal{Z}}  Z_\phi Y_\theta D(\alpha)e^{-\sum\limits_i\beta \omega_i n_i}D^\dagger(\alpha) Y_{-\theta}Z_{-\phi}
\end{eqnarray}
where $\mathcal{Z}=\Tr{\rho}=\prod_i(1-e^{-\beta\omega_i })^{-1}$ is the normalization factor, and ${D(\alpha) =e^{\alpha \hat{a}^\dagger _x - \alpha^{*} \hat{a}_x}}$ is the displacement operator.

The expectation value of the photon number in the $y$ cavity is given by
\begin{eqnarray}
    N_y = \Tr{\hat{N}_y \rho} = \kappa_{ext.}\tilde{N} h_2 + n_{th.}
\end{eqnarray}
where the trace is over all Fock states, $\tilde{N} = |\alpha|^2$ is the mean photon number of the coherent state, ${h_2 =g\left(\theta,\phi\right)\sin^2 \left(\frac{ |\bm B|}{2} t\right)} $ is a probability amplitude [c.f., Eq.~\eqref{eq:prob amp}], and $n_{th.} =  \kappa_{ext.}\mathcal{Z}e^{-\beta( \omega_x+ \omega_y)}$ is determined by the thermal distribution of the states.
Similarly, the variance of the photon number in the $y$ cavity is given by~\cite{xia_detection_2015}
\begin{eqnarray}
   \Delta^2 N_y =  \kappa_{ext.}^2\tilde{N} h_2(2 n_{th.} +  h_2) +  n_{th.} ^2 +  n_{th.}\,.
\end{eqnarray}
Assuming that the mean photon number of the input field is much larger than the mean photon number of thermal noise, i.e., $ \tilde N \gg n_{th.}$, the variance is approximated as $ \Delta^2 N_y \approx  \kappa_{ext.}\tilde{N} h_2(2 n_{th.} +  \kappa_{ext.}h_2)$

The precision in estimating the $y$ component of the vector $\bm B$, which is related to the Hall conductivity, is given by the mean-squared error
\begin{eqnarray}
\begin{array}{rl}
    \tau\Delta \bm B_y ^2& \gtrsim \tau_m\frac{ \Delta^2 N_y }{\left(\partial N_y / \partial \bm B_y\right)^2}\\
     &  \gtrsim \tau_m \frac{2 n_{th.} +h_2 }{\tilde{ N}F}
\end{array}
\end{eqnarray}
where $\tau$ is the coherence time of the device and ${\tau_m \sim |\bm B|^{-1}}$ is the duration of a single measurement which is determined by the period of oscillations.
The Fisher information is given by $F=\frac{1}{h_2}(\frac{\partial h_2}{\partial \bm B_y})^2$ and at the working time $t=\pi|\bm B|^{-1}$ it can be as high as $  F \sim |\bm B|^{-2}$.

\begin{acknowledgments}
\textbf{Acknowledgments}
This work is supported by the Quantum Science Center (QSC), a National Quantum Information Science Research Center of the U.S. Department of Energy (DOE). 
P.N. gratefully acknowledges support from the Gordon and Betty Moore Foundation grant No. 8048 and from the John Simon Guggenheim Memorial Foundation (Guggenheim Fellowship). 
A.Y. is also partly supported by the Gordon and Betty Moore Foundation through grant No. GBMF9468.
\end{acknowledgments}


\begin{thebibliography}{52}%
\makeatletter
\providecommand \@ifxundefined [1]{%
 \@ifx{#1\undefined}
}%
\providecommand \@ifnum [1]{%
 \ifnum #1\expandafter \@firstoftwo
 \else \expandafter \@secondoftwo
 \fi
}%
\providecommand \@ifx [1]{%
 \ifx #1\expandafter \@firstoftwo
 \else \expandafter \@secondoftwo
 \fi
}%
\providecommand \natexlab [1]{#1}%
\providecommand \enquote  [1]{``#1''}%
\providecommand \bibnamefont  [1]{#1}%
\providecommand \bibfnamefont [1]{#1}%
\providecommand \citenamefont [1]{#1}%
\providecommand \href@noop [0]{\@secondoftwo}%
\providecommand \href [0]{\begingroup \@sanitize@url \@href}%
\providecommand \@href[1]{\@@startlink{#1}\@@href}%
\providecommand \@@href[1]{\endgroup#1\@@endlink}%
\providecommand \@sanitize@url [0]{\catcode `\\12\catcode `\$12\catcode
  `\&12\catcode `\#12\catcode `\^12\catcode `\_12\catcode `\%12\relax}%
\providecommand \@@startlink[1]{}%
\providecommand \@@endlink[0]{}%
\providecommand \url  [0]{\begingroup\@sanitize@url \@url }%
\providecommand \@url [1]{\endgroup\@href {#1}{\urlprefix }}%
\providecommand \urlprefix  [0]{URL }%
\providecommand \Eprint [0]{\href }%
\providecommand \doibase [0]{https://doi.org/}%
\providecommand \selectlanguage [0]{\@gobble}%
\providecommand \bibinfo  [0]{\@secondoftwo}%
\providecommand \bibfield  [0]{\@secondoftwo}%
\providecommand \translation [1]{[#1]}%
\providecommand \BibitemOpen [0]{}%
\providecommand \bibitemStop [0]{}%
\providecommand \bibitemNoStop [0]{.\EOS\space}%
\providecommand \EOS [0]{\spacefactor3000\relax}%
\providecommand \BibitemShut  [1]{\csname bibitem#1\endcsname}%
\let\auto@bib@innerbib\@empty
\bibitem [{\citenamefont {Petrides}\ and\ \citenamefont
  {Zilberberg}(2022)}]{petrides2022semiclassical}%
  \BibitemOpen
  \bibfield  {author} {\bibinfo {author} {\bibfnamefont {I.}~\bibnamefont
  {Petrides}}\ and\ \bibinfo {author} {\bibfnamefont {O.}~\bibnamefont
  {Zilberberg}},\ }\bibfield  {title} {\bibinfo {title} {Semiclassical
  treatment of spinor topological effects in driven inhomogeneous insulators
  under external electromagnetic fields},\ }\href@noop {} {\bibfield  {journal}
  {\bibinfo  {journal} {Physical Review B}\ }\textbf {\bibinfo {volume}
  {106}},\ \bibinfo {pages} {165130} (\bibinfo {year} {2022})}\BibitemShut
  {NoStop}%
\bibitem [{\citenamefont {Zhao}\ \emph {et~al.}(2020)\citenamefont {Zhao},
  \citenamefont {Guo}, \citenamefont {Garcia}, \citenamefont {Narang},\ and\
  \citenamefont {Fan}}]{zhao2020axion}%
  \BibitemOpen
  \bibfield  {author} {\bibinfo {author} {\bibfnamefont {B.}~\bibnamefont
  {Zhao}}, \bibinfo {author} {\bibfnamefont {C.}~\bibnamefont {Guo}}, \bibinfo
  {author} {\bibfnamefont {C.~A.}\ \bibnamefont {Garcia}}, \bibinfo {author}
  {\bibfnamefont {P.}~\bibnamefont {Narang}},\ and\ \bibinfo {author}
  {\bibfnamefont {S.}~\bibnamefont {Fan}},\ }\bibfield  {title} {\bibinfo
  {title} {Axion-field-enabled nonreciprocal thermal radiation in weyl
  semimetals},\ }\href@noop {} {\bibfield  {journal} {\bibinfo  {journal} {Nano
  letters}\ }\textbf {\bibinfo {volume} {20}},\ \bibinfo {pages} {1923}
  (\bibinfo {year} {2020})}\BibitemShut {NoStop}%
\bibitem [{\citenamefont {Curtis}\ \emph
  {et~al.}(2023{\natexlab{a}})\citenamefont {Curtis}, \citenamefont
  {Petrides},\ and\ \citenamefont {Narang}}]{Curtis.2023a}%
  \BibitemOpen
  \bibfield  {author} {\bibinfo {author} {\bibfnamefont {J.}~\bibnamefont
  {Curtis}}, \bibinfo {author} {\bibfnamefont {I.}~\bibnamefont {Petrides}},\
  and\ \bibinfo {author} {\bibfnamefont {P.}~\bibnamefont {Narang}},\
  }\bibfield  {title} {\bibinfo {title} {Finite-momentum instability of
  dynamical axion insulator},\ }\href {https://doi.org/xxxxxx} {\bibfield
  {journal} {\bibinfo  {journal} {Phys. Rev. B}\ }\textbf {\bibinfo {volume}
  {xxx}},\ \bibinfo {pages} {xxxxx} (\bibinfo {year}
  {2023}{\natexlab{a}})}\BibitemShut {NoStop}%
\bibitem [{\citenamefont {Narang}\ \emph {et~al.}(2021)\citenamefont {Narang},
  \citenamefont {Garcia},\ and\ \citenamefont {Felser}}]{Narang.2021}%
  \BibitemOpen
  \bibfield  {author} {\bibinfo {author} {\bibfnamefont {P.}~\bibnamefont
  {Narang}}, \bibinfo {author} {\bibfnamefont {C.}~\bibnamefont {Garcia}},\
  and\ \bibinfo {author} {\bibfnamefont {C.}~\bibnamefont {Felser}},\
  }\bibfield  {title} {\bibinfo {title} {the topology of electronic band
  structures},\ }\href {https://doi.org/10.1038/s41563-020-00820-4} {\bibfield
  {journal} {\bibinfo  {journal} {Nature Mat.}\ }\textbf {\bibinfo {volume}
  {20}},\ \bibinfo {pages} {293} (\bibinfo {year} {2021})}\BibitemShut
  {NoStop}%
\bibitem [{\citenamefont {Nenno}\ \emph {et~al.}(2020)\citenamefont {Nenno},
  \citenamefont {Garcia}, \citenamefont {Gooth}, \citenamefont {Felser},\ and\
  \citenamefont {Narang}}]{Nenno.2020}%
  \BibitemOpen
  \bibfield  {author} {\bibinfo {author} {\bibfnamefont {D.}~\bibnamefont
  {Nenno}}, \bibinfo {author} {\bibfnamefont {C.}~\bibnamefont {Garcia}},
  \bibinfo {author} {\bibfnamefont {J.}~\bibnamefont {Gooth}}, \bibinfo
  {author} {\bibfnamefont {C.}~\bibnamefont {Felser}},\ and\ \bibinfo {author}
  {\bibfnamefont {P.}~\bibnamefont {Narang}},\ }\bibfield  {title} {\bibinfo
  {title} {Axion physics in condensed-matter systems},\ }\href
  {https://doi.org/10.1038/s42254-020-0240-2} {\bibfield  {journal} {\bibinfo
  {journal} {Nat. Rev. Phys.}\ }\textbf {\bibinfo {volume} {2}},\ \bibinfo
  {pages} {682} (\bibinfo {year} {2020})}\BibitemShut {NoStop}%
\bibitem [{\citenamefont {Vool}\ \emph {et~al.}(2021)\citenamefont {Vool},
  \citenamefont {Hamo}, \citenamefont {Varnavides}, \citenamefont {Wang},
  \citenamefont {Zhou}, \citenamefont {Kumar}, \citenamefont {Dovzhenko},
  \citenamefont {Qiu}, \citenamefont {Garcia}, \citenamefont {Pierce},
  \citenamefont {Gooth}, \citenamefont {Anikeeva}, \citenamefont {Felser},
  \citenamefont {Narang},\ and\ \citenamefont {Yacoby}}]{Vool.2021}%
  \BibitemOpen
  \bibfield  {author} {\bibinfo {author} {\bibfnamefont {U.}~\bibnamefont
  {Vool}}, \bibinfo {author} {\bibfnamefont {A.}~\bibnamefont {Hamo}}, \bibinfo
  {author} {\bibfnamefont {G.}~\bibnamefont {Varnavides}}, \bibinfo {author}
  {\bibfnamefont {Y.}~\bibnamefont {Wang}}, \bibinfo {author} {\bibfnamefont
  {T.}~\bibnamefont {Zhou}}, \bibinfo {author} {\bibfnamefont {N.}~\bibnamefont
  {Kumar}}, \bibinfo {author} {\bibfnamefont {Y.}~\bibnamefont {Dovzhenko}},
  \bibinfo {author} {\bibfnamefont {Z.}~\bibnamefont {Qiu}}, \bibinfo {author}
  {\bibfnamefont {C.}~\bibnamefont {Garcia}}, \bibinfo {author} {\bibfnamefont
  {A.}~\bibnamefont {Pierce}}, \bibinfo {author} {\bibfnamefont
  {J.}~\bibnamefont {Gooth}}, \bibinfo {author} {\bibfnamefont
  {P.}~\bibnamefont {Anikeeva}}, \bibinfo {author} {\bibfnamefont
  {C.}~\bibnamefont {Felser}}, \bibinfo {author} {\bibfnamefont
  {P.}~\bibnamefont {Narang}},\ and\ \bibinfo {author} {\bibfnamefont
  {A.}~\bibnamefont {Yacoby}},\ }\bibfield  {title} {\bibinfo {title} {Imaging
  phonon-mediated hydrodynamic flow in wte$_2$},\ }\href
  {https://doi.org/10.1038/s41567-021-01341-w} {\bibfield  {journal} {\bibinfo
  {journal} {Nature Phys.}\ }\textbf {\bibinfo {volume} {17}},\ \bibinfo
  {pages} {1216} (\bibinfo {year} {2021})}\BibitemShut {NoStop}%
\bibitem [{\citenamefont {Varnavides}\ \emph {et~al.}(2020)\citenamefont
  {Varnavides}, \citenamefont {Jermyn}, \citenamefont {Anikeeva}, \citenamefont
  {Fleser},\ and\ \citenamefont {Narang}}]{Varnavides.2020}%
  \BibitemOpen
  \bibfield  {author} {\bibinfo {author} {\bibfnamefont {G.}~\bibnamefont
  {Varnavides}}, \bibinfo {author} {\bibfnamefont {A.}~\bibnamefont {Jermyn}},
  \bibinfo {author} {\bibfnamefont {P.}~\bibnamefont {Anikeeva}}, \bibinfo
  {author} {\bibfnamefont {C.}~\bibnamefont {Fleser}},\ and\ \bibinfo {author}
  {\bibfnamefont {P.}~\bibnamefont {Narang}},\ }\bibfield  {title} {\bibinfo
  {title} {Electron hydrodynamics in anisotropic materials},\ }\href
  {https://doi.org/10.1038/s41467-020-18553-y} {\bibfield  {journal} {\bibinfo
  {journal} {Nat. Commun.}\ }\textbf {\bibinfo {volume} {11}},\ \bibinfo
  {pages} {4710} (\bibinfo {year} {2020})}\BibitemShut {NoStop}%
\bibitem [{\citenamefont {Basov}\ \emph {et~al.}(2011)\citenamefont {Basov},
  \citenamefont {Averitt}, \citenamefont {Van Der~Marel}, \citenamefont
  {Dressel},\ and\ \citenamefont {Haule}}]{basov2011electrodynamics}%
  \BibitemOpen
  \bibfield  {author} {\bibinfo {author} {\bibfnamefont {D.~N.}\ \bibnamefont
  {Basov}}, \bibinfo {author} {\bibfnamefont {R.~D.}\ \bibnamefont {Averitt}},
  \bibinfo {author} {\bibfnamefont {D.}~\bibnamefont {Van Der~Marel}}, \bibinfo
  {author} {\bibfnamefont {M.}~\bibnamefont {Dressel}},\ and\ \bibinfo {author}
  {\bibfnamefont {K.}~\bibnamefont {Haule}},\ }\bibfield  {title} {\bibinfo
  {title} {Electrodynamics of correlated electron materials},\ }\href@noop {}
  {\bibfield  {journal} {\bibinfo  {journal} {Reviews of Modern Physics}\
  }\textbf {\bibinfo {volume} {83}},\ \bibinfo {pages} {471} (\bibinfo {year}
  {2011})}\BibitemShut {NoStop}%
\bibitem [{\citenamefont {Potton}(2004)}]{potton2004reciprocity}%
  \BibitemOpen
  \bibfield  {author} {\bibinfo {author} {\bibfnamefont {R.~J.}\ \bibnamefont
  {Potton}},\ }\bibfield  {title} {\bibinfo {title} {Reciprocity in optics},\
  }\href@noop {} {\bibfield  {journal} {\bibinfo  {journal} {Reports on
  Progress in Physics}\ }\textbf {\bibinfo {volume} {67}},\ \bibinfo {pages}
  {717} (\bibinfo {year} {2004})}\BibitemShut {NoStop}%
\bibitem [{\citenamefont {Asadchy}\ \emph {et~al.}(2020)\citenamefont
  {Asadchy}, \citenamefont {Mirmoosa}, \citenamefont {Diaz-Rubio},
  \citenamefont {Fan},\ and\ \citenamefont {Tretyakov}}]{asadchy2020tutorial}%
  \BibitemOpen
  \bibfield  {author} {\bibinfo {author} {\bibfnamefont {V.~S.}\ \bibnamefont
  {Asadchy}}, \bibinfo {author} {\bibfnamefont {M.~S.}\ \bibnamefont
  {Mirmoosa}}, \bibinfo {author} {\bibfnamefont {A.}~\bibnamefont
  {Diaz-Rubio}}, \bibinfo {author} {\bibfnamefont {S.}~\bibnamefont {Fan}},\
  and\ \bibinfo {author} {\bibfnamefont {S.~A.}\ \bibnamefont {Tretyakov}},\
  }\bibfield  {title} {\bibinfo {title} {Tutorial on electromagnetic
  nonreciprocity and its origins},\ }\href@noop {} {\bibfield  {journal}
  {\bibinfo  {journal} {Proceedings of the IEEE}\ }\textbf {\bibinfo {volume}
  {108}},\ \bibinfo {pages} {1684} (\bibinfo {year} {2020})}\BibitemShut
  {NoStop}%
\bibitem [{\citenamefont {Yu}\ \emph {et~al.}(2010)\citenamefont {Yu},
  \citenamefont {Zhang}, \citenamefont {Zhang}, \citenamefont {Zhang},
  \citenamefont {Dai},\ and\ \citenamefont {Fang}}]{yu_quantized_2010}%
  \BibitemOpen
  \bibfield  {author} {\bibinfo {author} {\bibfnamefont {R.}~\bibnamefont
  {Yu}}, \bibinfo {author} {\bibfnamefont {W.}~\bibnamefont {Zhang}}, \bibinfo
  {author} {\bibfnamefont {H.-J.}\ \bibnamefont {Zhang}}, \bibinfo {author}
  {\bibfnamefont {S.-C.}\ \bibnamefont {Zhang}}, \bibinfo {author}
  {\bibfnamefont {X.}~\bibnamefont {Dai}},\ and\ \bibinfo {author}
  {\bibfnamefont {Z.}~\bibnamefont {Fang}},\ }\bibfield  {title} {\bibinfo
  {title} {Quantized {Anomalous} {Hall} {Effect} in {Magnetic} {Topological}
  {Insulators}},\ }\href {https://doi.org/10.1126/science.1187485} {\bibfield
  {journal} {\bibinfo  {journal} {Science}\ }\textbf {\bibinfo {volume}
  {329}},\ \bibinfo {pages} {61} (\bibinfo {year} {2010})}\BibitemShut
  {NoStop}%
\bibitem [{\citenamefont {da~Silva~Neto}(2019)}]{da_silva_neto_weyling_2019}%
  \BibitemOpen
  \bibfield  {author} {\bibinfo {author} {\bibfnamefont {E.~H.}\ \bibnamefont
  {da~Silva~Neto}},\ }\bibfield  {title} {\bibinfo {title} {“{Weyl}”ing
  away time-reversal symmetry},\ }\href
  {https://doi.org/10.1126/science.aax6190} {\bibfield  {journal} {\bibinfo
  {journal} {Science}\ }\textbf {\bibinfo {volume} {365}},\ \bibinfo {pages}
  {1248} (\bibinfo {year} {2019})}\BibitemShut {NoStop}%
\bibitem [{\citenamefont {Kuiri}\ \emph {et~al.}(2022)\citenamefont {Kuiri},
  \citenamefont {Coleman}, \citenamefont {Gao}, \citenamefont {Vishnuradhan},
  \citenamefont {Watanabe}, \citenamefont {Taniguchi}, \citenamefont {Zhu},
  \citenamefont {MacDonald},\ and\ \citenamefont
  {Folk}}]{corr_kuiri_spontaneous_2022}%
  \BibitemOpen
  \bibfield  {author} {\bibinfo {author} {\bibfnamefont {M.}~\bibnamefont
  {Kuiri}}, \bibinfo {author} {\bibfnamefont {C.}~\bibnamefont {Coleman}},
  \bibinfo {author} {\bibfnamefont {Z.}~\bibnamefont {Gao}}, \bibinfo {author}
  {\bibfnamefont {A.}~\bibnamefont {Vishnuradhan}}, \bibinfo {author}
  {\bibfnamefont {K.}~\bibnamefont {Watanabe}}, \bibinfo {author}
  {\bibfnamefont {T.}~\bibnamefont {Taniguchi}}, \bibinfo {author}
  {\bibfnamefont {J.}~\bibnamefont {Zhu}}, \bibinfo {author} {\bibfnamefont
  {A.~H.}\ \bibnamefont {MacDonald}},\ and\ \bibinfo {author} {\bibfnamefont
  {J.}~\bibnamefont {Folk}},\ }\bibfield  {title} {\bibinfo {title}
  {Spontaneous time-reversal symmetry breaking in twisted double bilayer
  graphene},\ }\href {https://doi.org/10.1038/s41467-022-34192-x} {\bibfield
  {journal} {\bibinfo  {journal} {Nature Communications}\ }\textbf {\bibinfo
  {volume} {13}},\ \bibinfo {pages} {6468} (\bibinfo {year}
  {2022})}\BibitemShut {NoStop}%
\bibitem [{\citenamefont {Lee}\ \emph {et~al.}(2019)\citenamefont {Lee},
  \citenamefont {Khalaf}, \citenamefont {Liu}, \citenamefont {Liu},
  \citenamefont {Hao}, \citenamefont {Kim},\ and\ \citenamefont
  {Vishwanath}}]{corr_lee_theory_2019}%
  \BibitemOpen
  \bibfield  {author} {\bibinfo {author} {\bibfnamefont {J.~Y.}\ \bibnamefont
  {Lee}}, \bibinfo {author} {\bibfnamefont {E.}~\bibnamefont {Khalaf}},
  \bibinfo {author} {\bibfnamefont {S.}~\bibnamefont {Liu}}, \bibinfo {author}
  {\bibfnamefont {X.}~\bibnamefont {Liu}}, \bibinfo {author} {\bibfnamefont
  {Z.}~\bibnamefont {Hao}}, \bibinfo {author} {\bibfnamefont {P.}~\bibnamefont
  {Kim}},\ and\ \bibinfo {author} {\bibfnamefont {A.}~\bibnamefont
  {Vishwanath}},\ }\bibfield  {title} {\bibinfo {title} {Theory of correlated
  insulating behaviour and spin-triplet superconductivity in twisted double
  bilayer graphene},\ }\href {https://doi.org/10.1038/s41467-019-12981-1}
  {\bibfield  {journal} {\bibinfo  {journal} {Nature Communications}\ }\textbf
  {\bibinfo {volume} {10}},\ \bibinfo {pages} {5333} (\bibinfo {year}
  {2019})}\BibitemShut {NoStop}%
\bibitem [{\citenamefont {Liu}\ \emph {et~al.}(2020)\citenamefont {Liu},
  \citenamefont {Hao}, \citenamefont {Khalaf}, \citenamefont {Lee},
  \citenamefont {Ronen}, \citenamefont {Yoo}, \citenamefont {Haei~Najafabadi},
  \citenamefont {Watanabe}, \citenamefont {Taniguchi}, \citenamefont
  {Vishwanath},\ and\ \citenamefont {Kim}}]{corr_liu_tunable_2020}%
  \BibitemOpen
  \bibfield  {author} {\bibinfo {author} {\bibfnamefont {X.}~\bibnamefont
  {Liu}}, \bibinfo {author} {\bibfnamefont {Z.}~\bibnamefont {Hao}}, \bibinfo
  {author} {\bibfnamefont {E.}~\bibnamefont {Khalaf}}, \bibinfo {author}
  {\bibfnamefont {J.~Y.}\ \bibnamefont {Lee}}, \bibinfo {author} {\bibfnamefont
  {Y.}~\bibnamefont {Ronen}}, \bibinfo {author} {\bibfnamefont
  {H.}~\bibnamefont {Yoo}}, \bibinfo {author} {\bibfnamefont {D.}~\bibnamefont
  {Haei~Najafabadi}}, \bibinfo {author} {\bibfnamefont {K.}~\bibnamefont
  {Watanabe}}, \bibinfo {author} {\bibfnamefont {T.}~\bibnamefont {Taniguchi}},
  \bibinfo {author} {\bibfnamefont {A.}~\bibnamefont {Vishwanath}},\ and\
  \bibinfo {author} {\bibfnamefont {P.}~\bibnamefont {Kim}},\ }\bibfield
  {title} {\bibinfo {title} {Tunable spin-polarized correlated states in
  twisted double bilayer graphene},\ }\href
  {https://doi.org/10.1038/s41586-020-2458-7} {\bibfield  {journal} {\bibinfo
  {journal} {Nature}\ }\textbf {\bibinfo {volume} {583}},\ \bibinfo {pages}
  {221} (\bibinfo {year} {2020})}\BibitemShut {NoStop}%
\bibitem [{\citenamefont {Ghosh}\ \emph {et~al.}(2020)\citenamefont {Ghosh},
  \citenamefont {Smidman}, \citenamefont {Shang}, \citenamefont {Annett},
  \citenamefont {Hillier}, \citenamefont {Quintanilla},\ and\ \citenamefont
  {Yuan}}]{ghosh2020recent}%
  \BibitemOpen
  \bibfield  {author} {\bibinfo {author} {\bibfnamefont {S.~K.}\ \bibnamefont
  {Ghosh}}, \bibinfo {author} {\bibfnamefont {M.}~\bibnamefont {Smidman}},
  \bibinfo {author} {\bibfnamefont {T.}~\bibnamefont {Shang}}, \bibinfo
  {author} {\bibfnamefont {J.~F.}\ \bibnamefont {Annett}}, \bibinfo {author}
  {\bibfnamefont {A.~D.}\ \bibnamefont {Hillier}}, \bibinfo {author}
  {\bibfnamefont {J.}~\bibnamefont {Quintanilla}},\ and\ \bibinfo {author}
  {\bibfnamefont {H.}~\bibnamefont {Yuan}},\ }\bibfield  {title} {\bibinfo
  {title} {Recent progress on superconductors with time-reversal symmetry
  breaking},\ }\href@noop {} {\bibfield  {journal} {\bibinfo  {journal}
  {Journal of Physics: Condensed Matter}\ }\textbf {\bibinfo {volume} {33}},\
  \bibinfo {pages} {033001} (\bibinfo {year} {2020})}\BibitemShut {NoStop}%
\bibitem [{\citenamefont {Read}\ and\ \citenamefont {Green}(2000)}]{Read.2000}%
  \BibitemOpen
  \bibfield  {author} {\bibinfo {author} {\bibfnamefont {N.}~\bibnamefont
  {Read}}\ and\ \bibinfo {author} {\bibfnamefont {D.}~\bibnamefont {Green}},\
  }\bibfield  {title} {\bibinfo {title} {{Paired states of fermions in two
  dimensions with breaking of parity and time-reversal symmetries and the
  fractional quantum Hall effect}},\ }\href
  {https://doi.org/10.1103/physrevb.61.10267} {\bibfield  {journal} {\bibinfo
  {journal} {Physical Review B}\ }\textbf {\bibinfo {volume} {61}},\ \bibinfo
  {pages} {10267} (\bibinfo {year} {2000})}\BibitemShut {NoStop}%
\bibitem [{\citenamefont {Kallin}\ and\ \citenamefont
  {Berlinsky}(2016)}]{kallin_chiral_2016}%
  \BibitemOpen
  \bibfield  {author} {\bibinfo {author} {\bibfnamefont {C.}~\bibnamefont
  {Kallin}}\ and\ \bibinfo {author} {\bibfnamefont {J.}~\bibnamefont
  {Berlinsky}},\ }\bibfield  {title} {\bibinfo {title} {Chiral
  superconductors},\ }\href {https://doi.org/10.1088/0034-4885/79/5/054502}
  {\bibfield  {journal} {\bibinfo  {journal} {Reports on Progress in Physics}\
  }\textbf {\bibinfo {volume} {79}},\ \bibinfo {pages} {054502} (\bibinfo
  {year} {2016})}\BibitemShut {NoStop}%
\bibitem [{\citenamefont {Poniatowski}\ \emph {et~al.}(2021)\citenamefont
  {Poniatowski}, \citenamefont {Curtis}, \citenamefont {Yacoby},\ and\
  \citenamefont {Narang}}]{Poniatowski.2022a}%
  \BibitemOpen
  \bibfield  {author} {\bibinfo {author} {\bibfnamefont {N.}~\bibnamefont
  {Poniatowski}}, \bibinfo {author} {\bibfnamefont {J.}~\bibnamefont {Curtis}},
  \bibinfo {author} {\bibfnamefont {A.}~\bibnamefont {Yacoby}},\ and\ \bibinfo
  {author} {\bibfnamefont {P.}~\bibnamefont {Narang}},\ }\bibfield  {title}
  {\bibinfo {title} {Spectroscopic signatures of time-reversal symmetry
  breaking superconductivity},\ }\href
  {https://doi.org/10.1038/s42005-022-00819-0} {\bibfield  {journal} {\bibinfo
  {journal} {Comm. Phys.}\ }\textbf {\bibinfo {volume} {5}},\ \bibinfo {pages}
  {44} (\bibinfo {year} {2021})}\BibitemShut {NoStop}%
\bibitem [{\citenamefont {Curtis}\ \emph {et~al.}(2022)\citenamefont {Curtis},
  \citenamefont {Poniatowski}, \citenamefont {Yacoby},\ and\ \citenamefont
  {Narang}}]{Curtis.2022a}%
  \BibitemOpen
  \bibfield  {author} {\bibinfo {author} {\bibfnamefont {J.}~\bibnamefont
  {Curtis}}, \bibinfo {author} {\bibfnamefont {N.}~\bibnamefont {Poniatowski}},
  \bibinfo {author} {\bibfnamefont {A.}~\bibnamefont {Yacoby}},\ and\ \bibinfo
  {author} {\bibfnamefont {P.}~\bibnamefont {Narang}},\ }\bibfield  {title}
  {\bibinfo {title} {Proximity-induced collective modes in an unconventional
  superconductor heterostructure},\ }\href
  {https://doi.org/10.1103/PhysRevB.106.064508} {\bibfield  {journal} {\bibinfo
   {journal} {Phys. Rev. B}\ }\textbf {\bibinfo {volume} {106}},\ \bibinfo
  {pages} {064508} (\bibinfo {year} {2022})}\BibitemShut {NoStop}%
\bibitem [{\citenamefont {Poniatowski}\ \emph {et~al.}(2022)\citenamefont
  {Poniatowski}, \citenamefont {Curtis}, \citenamefont {B{\o}ttcher},
  \citenamefont {Galitski}, \citenamefont {Yacoby}, \citenamefont {Narang},\
  and\ \citenamefont {Demler}}]{Poniatowski.2022b}%
  \BibitemOpen
  \bibfield  {author} {\bibinfo {author} {\bibfnamefont {N.}~\bibnamefont
  {Poniatowski}}, \bibinfo {author} {\bibfnamefont {J.}~\bibnamefont {Curtis}},
  \bibinfo {author} {\bibfnamefont {C.}~\bibnamefont {B{\o}ttcher}}, \bibinfo
  {author} {\bibfnamefont {V.}~\bibnamefont {Galitski}}, \bibinfo {author}
  {\bibfnamefont {A.}~\bibnamefont {Yacoby}}, \bibinfo {author} {\bibfnamefont
  {P.}~\bibnamefont {Narang}},\ and\ \bibinfo {author} {\bibfnamefont
  {E.}~\bibnamefont {Demler}},\ }\bibfield  {title} {\bibinfo {title} {Surface
  cooper-pair spin waves in triplet superconductors},\ }\href
  {https://doi.org/10.1103/PhysRevLett.129.237002} {\bibfield  {journal}
  {\bibinfo  {journal} {Phys. Rev. Lett.}\ }\textbf {\bibinfo {volume} {129}},\
  \bibinfo {pages} {237002} (\bibinfo {year} {2022})}\BibitemShut {NoStop}%
\bibitem [{\citenamefont {Curtis}\ \emph
  {et~al.}(2023{\natexlab{b}})\citenamefont {Curtis}, \citenamefont
  {Poniatowski}, \citenamefont {Xie}, \citenamefont {Yacoby}, \citenamefont
  {Demler},\ and\ \citenamefont {Narang}}]{Curtis.2023b}%
  \BibitemOpen
  \bibfield  {author} {\bibinfo {author} {\bibfnamefont {J.}~\bibnamefont
  {Curtis}}, \bibinfo {author} {\bibfnamefont {N.}~\bibnamefont {Poniatowski}},
  \bibinfo {author} {\bibfnamefont {Y.}~\bibnamefont {Xie}}, \bibinfo {author}
  {\bibfnamefont {A.}~\bibnamefont {Yacoby}}, \bibinfo {author} {\bibfnamefont
  {E.}~\bibnamefont {Demler}},\ and\ \bibinfo {author} {\bibfnamefont
  {P.}~\bibnamefont {Narang}},\ }\bibfield  {title} {\bibinfo {title}
  {Stabilizing fluctuating spin-triplet superconductivity in graphene via
  induced spin-orbit coupling},\ }\href
  {https://doi.org/10.1103/PhysRevLett.xxx.xxxxx} {\bibfield  {journal}
  {\bibinfo  {journal} {Phys. Rev. Lett.}\ }\textbf {\bibinfo {volume} {xxx}},\
  \bibinfo {pages} {xxxxxx} (\bibinfo {year} {2023}{\natexlab{b}})}\BibitemShut
  {NoStop}%
\bibitem [{\citenamefont {Xia}\ \emph {et~al.}(2006)\citenamefont {Xia},
  \citenamefont {Maeno}, \citenamefont {Beyersdorf}, \citenamefont {Fejer},\
  and\ \citenamefont {Kapitulnik}}]{xia_high_2006}%
  \BibitemOpen
  \bibfield  {author} {\bibinfo {author} {\bibfnamefont {J.}~\bibnamefont
  {Xia}}, \bibinfo {author} {\bibfnamefont {Y.}~\bibnamefont {Maeno}}, \bibinfo
  {author} {\bibfnamefont {P.~T.}\ \bibnamefont {Beyersdorf}}, \bibinfo
  {author} {\bibfnamefont {M.~M.}\ \bibnamefont {Fejer}},\ and\ \bibinfo
  {author} {\bibfnamefont {A.}~\bibnamefont {Kapitulnik}},\ }\bibfield  {title}
  {\bibinfo {title} {High {Resolution} {Polar} {Kerr} {Effect} {Measurements}
  of {Sr} 2{RuO}4 : {Evidence} for {Broken} {Time}-{Reversal} {Symmetry} in the
  {Superconducting} {State}},\ }\href@noop {} {\bibfield  {journal} {\bibinfo
  {journal} {Physical Review Letters}\ }\textbf {\bibinfo {volume} {97}},\
  \bibinfo {pages} {167002} (\bibinfo {year} {2006})}\BibitemShut {NoStop}%
\bibitem [{\citenamefont {Grinenko}\ \emph {et~al.}(2020)\citenamefont
  {Grinenko}, \citenamefont {Sarkar}, \citenamefont {Kihou}, \citenamefont
  {Lee}, \citenamefont {Morozov}, \citenamefont {Aswartham}, \citenamefont
  {Büchner}, \citenamefont {Chekhonin}, \citenamefont {Skrotzki},
  \citenamefont {Nenkov}, \citenamefont {Hühne}, \citenamefont {Nielsch},
  \citenamefont {Drechsler}, \citenamefont {Vadimov}, \citenamefont {Silaev},
  \citenamefont {Volkov}, \citenamefont {Eremin}, \citenamefont {Luetkens},\
  and\ \citenamefont {Klauss}}]{grinenko_superconductivity_2020}%
  \BibitemOpen
  \bibfield  {author} {\bibinfo {author} {\bibfnamefont {V.}~\bibnamefont
  {Grinenko}}, \bibinfo {author} {\bibfnamefont {R.}~\bibnamefont {Sarkar}},
  \bibinfo {author} {\bibfnamefont {K.}~\bibnamefont {Kihou}}, \bibinfo
  {author} {\bibfnamefont {C.~H.}\ \bibnamefont {Lee}}, \bibinfo {author}
  {\bibfnamefont {I.}~\bibnamefont {Morozov}}, \bibinfo {author} {\bibfnamefont
  {S.}~\bibnamefont {Aswartham}}, \bibinfo {author} {\bibfnamefont
  {B.}~\bibnamefont {Büchner}}, \bibinfo {author} {\bibfnamefont
  {P.}~\bibnamefont {Chekhonin}}, \bibinfo {author} {\bibfnamefont
  {W.}~\bibnamefont {Skrotzki}}, \bibinfo {author} {\bibfnamefont
  {K.}~\bibnamefont {Nenkov}}, \bibinfo {author} {\bibfnamefont
  {R.}~\bibnamefont {Hühne}}, \bibinfo {author} {\bibfnamefont
  {K.}~\bibnamefont {Nielsch}}, \bibinfo {author} {\bibfnamefont {S.~L.}\
  \bibnamefont {Drechsler}}, \bibinfo {author} {\bibfnamefont {V.~L.}\
  \bibnamefont {Vadimov}}, \bibinfo {author} {\bibfnamefont {M.~A.}\
  \bibnamefont {Silaev}}, \bibinfo {author} {\bibfnamefont {P.~A.}\
  \bibnamefont {Volkov}}, \bibinfo {author} {\bibfnamefont {I.}~\bibnamefont
  {Eremin}}, \bibinfo {author} {\bibfnamefont {H.}~\bibnamefont {Luetkens}},\
  and\ \bibinfo {author} {\bibfnamefont {H.-H.}\ \bibnamefont {Klauss}},\
  }\bibfield  {title} {\bibinfo {title} {Superconductivity with broken
  time-reversal symmetry inside a superconducting s-wave state},\ }\href
  {https://doi.org/10.1038/s41567-020-0886-9} {\bibfield  {journal} {\bibinfo
  {journal} {Nature Physics}\ }\textbf {\bibinfo {volume} {16}},\ \bibinfo
  {pages} {789} (\bibinfo {year} {2020})}\BibitemShut {NoStop}%
\bibitem [{\citenamefont {Mielke}\ \emph {et~al.}(2022)\citenamefont {Mielke},
  \citenamefont {Das}, \citenamefont {Yin}, \citenamefont {Liu}, \citenamefont
  {Gupta}, \citenamefont {Jiang}, \citenamefont {Medarde}, \citenamefont {Wu},
  \citenamefont {Lei}, \citenamefont {Chang}, \citenamefont {Dai},
  \citenamefont {Si}, \citenamefont {Miao}, \citenamefont {Thomale},
  \citenamefont {Neupert}, \citenamefont {Shi}, \citenamefont {Khasanov},
  \citenamefont {Hasan}, \citenamefont {Luetkens},\ and\ \citenamefont
  {Guguchia}}]{mielke_time-reversal_2022}%
  \BibitemOpen
  \bibfield  {author} {\bibinfo {author} {\bibfnamefont {C.}~\bibnamefont
  {Mielke}}, \bibinfo {author} {\bibfnamefont {D.}~\bibnamefont {Das}},
  \bibinfo {author} {\bibfnamefont {J.-X.}\ \bibnamefont {Yin}}, \bibinfo
  {author} {\bibfnamefont {H.}~\bibnamefont {Liu}}, \bibinfo {author}
  {\bibfnamefont {R.}~\bibnamefont {Gupta}}, \bibinfo {author} {\bibfnamefont
  {Y.-X.}\ \bibnamefont {Jiang}}, \bibinfo {author} {\bibfnamefont
  {M.}~\bibnamefont {Medarde}}, \bibinfo {author} {\bibfnamefont
  {X.}~\bibnamefont {Wu}}, \bibinfo {author} {\bibfnamefont {H.~C.}\
  \bibnamefont {Lei}}, \bibinfo {author} {\bibfnamefont {J.}~\bibnamefont
  {Chang}}, \bibinfo {author} {\bibfnamefont {P.}~\bibnamefont {Dai}}, \bibinfo
  {author} {\bibfnamefont {Q.}~\bibnamefont {Si}}, \bibinfo {author}
  {\bibfnamefont {H.}~\bibnamefont {Miao}}, \bibinfo {author} {\bibfnamefont
  {R.}~\bibnamefont {Thomale}}, \bibinfo {author} {\bibfnamefont
  {T.}~\bibnamefont {Neupert}}, \bibinfo {author} {\bibfnamefont
  {Y.}~\bibnamefont {Shi}}, \bibinfo {author} {\bibfnamefont {R.}~\bibnamefont
  {Khasanov}}, \bibinfo {author} {\bibfnamefont {M.~Z.}\ \bibnamefont {Hasan}},
  \bibinfo {author} {\bibfnamefont {H.}~\bibnamefont {Luetkens}},\ and\
  \bibinfo {author} {\bibfnamefont {Z.}~\bibnamefont {Guguchia}},\ }\bibfield
  {title} {\bibinfo {title} {Time-reversal symmetry-breaking charge order in a
  kagome superconductor},\ }\href {https://doi.org/10.1038/s41586-021-04327-z}
  {\bibfield  {journal} {\bibinfo  {journal} {Nature}\ }\textbf {\bibinfo
  {volume} {602}},\ \bibinfo {pages} {245} (\bibinfo {year}
  {2022})}\BibitemShut {NoStop}%
\bibitem [{\citenamefont {Grinenko}\ \emph {et~al.}(2021)\citenamefont
  {Grinenko}, \citenamefont {Ghosh}, \citenamefont {Sarkar}, \citenamefont
  {Orain}, \citenamefont {Nikitin}, \citenamefont {Elender}, \citenamefont
  {Das}, \citenamefont {Guguchia}, \citenamefont {Brückner}, \citenamefont
  {Barber}, \citenamefont {Park}, \citenamefont {Kikugawa}, \citenamefont
  {Sokolov}, \citenamefont {Bobowski}, \citenamefont {Miyoshi}, \citenamefont
  {Maeno}, \citenamefont {Mackenzie}, \citenamefont {Luetkens}, \citenamefont
  {Hicks},\ and\ \citenamefont {Klauss}}]{grinenko_split_2021}%
  \BibitemOpen
  \bibfield  {author} {\bibinfo {author} {\bibfnamefont {V.}~\bibnamefont
  {Grinenko}}, \bibinfo {author} {\bibfnamefont {S.}~\bibnamefont {Ghosh}},
  \bibinfo {author} {\bibfnamefont {R.}~\bibnamefont {Sarkar}}, \bibinfo
  {author} {\bibfnamefont {J.-C.}\ \bibnamefont {Orain}}, \bibinfo {author}
  {\bibfnamefont {A.}~\bibnamefont {Nikitin}}, \bibinfo {author} {\bibfnamefont
  {M.}~\bibnamefont {Elender}}, \bibinfo {author} {\bibfnamefont
  {D.}~\bibnamefont {Das}}, \bibinfo {author} {\bibfnamefont {Z.}~\bibnamefont
  {Guguchia}}, \bibinfo {author} {\bibfnamefont {F.}~\bibnamefont {Brückner}},
  \bibinfo {author} {\bibfnamefont {M.~E.}\ \bibnamefont {Barber}}, \bibinfo
  {author} {\bibfnamefont {J.}~\bibnamefont {Park}}, \bibinfo {author}
  {\bibfnamefont {N.}~\bibnamefont {Kikugawa}}, \bibinfo {author}
  {\bibfnamefont {D.~A.}\ \bibnamefont {Sokolov}}, \bibinfo {author}
  {\bibfnamefont {J.~S.}\ \bibnamefont {Bobowski}}, \bibinfo {author}
  {\bibfnamefont {T.}~\bibnamefont {Miyoshi}}, \bibinfo {author} {\bibfnamefont
  {Y.}~\bibnamefont {Maeno}}, \bibinfo {author} {\bibfnamefont {A.~P.}\
  \bibnamefont {Mackenzie}}, \bibinfo {author} {\bibfnamefont {H.}~\bibnamefont
  {Luetkens}}, \bibinfo {author} {\bibfnamefont {C.~W.}\ \bibnamefont
  {Hicks}},\ and\ \bibinfo {author} {\bibfnamefont {H.-H.}\ \bibnamefont
  {Klauss}},\ }\bibfield  {title} {\bibinfo {title} {Split superconducting and
  time-reversal symmetry-breaking transitions in {Sr2RuO4} under stress},\
  }\href {https://doi.org/10.1038/s41567-021-01182-7} {\bibfield  {journal}
  {\bibinfo  {journal} {Nature Physics}\ }\textbf {\bibinfo {volume} {17}},\
  \bibinfo {pages} {748} (\bibinfo {year} {2021})}\BibitemShut {NoStop}%
\bibitem [{\citenamefont {Wei}\ \emph {et~al.}(2022)\citenamefont {Wei},
  \citenamefont {Saykin}, \citenamefont {Miller}, \citenamefont {Ran},
  \citenamefont {Saha}, \citenamefont {Agterberg}, \citenamefont {Schmalian},
  \citenamefont {Butch}, \citenamefont {Paglione},\ and\ \citenamefont
  {Kapitulnik}}]{wei_interplay_2022}%
  \BibitemOpen
  \bibfield  {author} {\bibinfo {author} {\bibfnamefont {D.~S.}\ \bibnamefont
  {Wei}}, \bibinfo {author} {\bibfnamefont {D.}~\bibnamefont {Saykin}},
  \bibinfo {author} {\bibfnamefont {O.~Y.}\ \bibnamefont {Miller}}, \bibinfo
  {author} {\bibfnamefont {S.}~\bibnamefont {Ran}}, \bibinfo {author}
  {\bibfnamefont {S.~R.}\ \bibnamefont {Saha}}, \bibinfo {author}
  {\bibfnamefont {D.~F.}\ \bibnamefont {Agterberg}}, \bibinfo {author}
  {\bibfnamefont {J.}~\bibnamefont {Schmalian}}, \bibinfo {author}
  {\bibfnamefont {N.~P.}\ \bibnamefont {Butch}}, \bibinfo {author}
  {\bibfnamefont {J.}~\bibnamefont {Paglione}},\ and\ \bibinfo {author}
  {\bibfnamefont {A.}~\bibnamefont {Kapitulnik}},\ }\bibfield  {title}
  {\bibinfo {title} {Interplay between magnetism and superconductivity in {UTe}
  2},\ }\href {https://doi.org/10.1103/PhysRevB.105.024521} {\bibfield
  {journal} {\bibinfo  {journal} {Physical Review B}\ }\textbf {\bibinfo
  {volume} {105}},\ \bibinfo {pages} {024521} (\bibinfo {year}
  {2022})}\BibitemShut {NoStop}%
\bibitem [{\citenamefont {Schemm}\ \emph {et~al.}(2014)\citenamefont {Schemm},
  \citenamefont {Gannon}, \citenamefont {Wishne}, \citenamefont {Halperin},\
  and\ \citenamefont {Kapitulnik}}]{schemm_observation_2014}%
  \BibitemOpen
  \bibfield  {author} {\bibinfo {author} {\bibfnamefont {E.~R.}\ \bibnamefont
  {Schemm}}, \bibinfo {author} {\bibfnamefont {W.~J.}\ \bibnamefont {Gannon}},
  \bibinfo {author} {\bibfnamefont {C.~M.}\ \bibnamefont {Wishne}}, \bibinfo
  {author} {\bibfnamefont {W.~P.}\ \bibnamefont {Halperin}},\ and\ \bibinfo
  {author} {\bibfnamefont {A.}~\bibnamefont {Kapitulnik}},\ }\bibfield  {title}
  {\bibinfo {title} {Observation of broken time-reversal symmetry in the
  heavy-fermion superconductor {UPt} $_{\textrm{3}}$},\ }\href
  {https://doi.org/10.1126/science.1248552} {\bibfield  {journal} {\bibinfo
  {journal} {Science}\ }\textbf {\bibinfo {volume} {345}},\ \bibinfo {pages}
  {190} (\bibinfo {year} {2014})}\BibitemShut {NoStop}%
\bibitem [{\citenamefont {Gong}\ \emph {et~al.}(2017)\citenamefont {Gong},
  \citenamefont {Kargarian}, \citenamefont {Stern}, \citenamefont {Yue},
  \citenamefont {Zhou}, \citenamefont {Jin}, \citenamefont {Galitski},
  \citenamefont {Yakovenko},\ and\ \citenamefont
  {Xia}}]{gong_time-reversal_2017}%
  \BibitemOpen
  \bibfield  {author} {\bibinfo {author} {\bibfnamefont {X.}~\bibnamefont
  {Gong}}, \bibinfo {author} {\bibfnamefont {M.}~\bibnamefont {Kargarian}},
  \bibinfo {author} {\bibfnamefont {A.}~\bibnamefont {Stern}}, \bibinfo
  {author} {\bibfnamefont {D.}~\bibnamefont {Yue}}, \bibinfo {author}
  {\bibfnamefont {H.}~\bibnamefont {Zhou}}, \bibinfo {author} {\bibfnamefont
  {X.}~\bibnamefont {Jin}}, \bibinfo {author} {\bibfnamefont {V.~M.}\
  \bibnamefont {Galitski}}, \bibinfo {author} {\bibfnamefont {V.~M.}\
  \bibnamefont {Yakovenko}},\ and\ \bibinfo {author} {\bibfnamefont
  {J.}~\bibnamefont {Xia}},\ }\bibfield  {title} {\bibinfo {title}
  {Time-reversal symmetry-breaking superconductivity in epitaxial
  bismuth/nickel bilayers},\ }\href {https://doi.org/10.1126/sciadv.1602579}
  {\bibfield  {journal} {\bibinfo  {journal} {Science Advances}\ }\textbf
  {\bibinfo {volume} {3}},\ \bibinfo {pages} {e1602579} (\bibinfo {year}
  {2017})}\BibitemShut {NoStop}%
\bibitem [{\citenamefont {Cao}\ \emph {et~al.}(2018)\citenamefont {Cao},
  \citenamefont {Fatemi}, \citenamefont {Fang}, \citenamefont {Watanabe},
  \citenamefont {Taniguchi}, \citenamefont {Kaxiras},\ and\ \citenamefont
  {Jarillo-Herrero}}]{cao_unconventional_2018}%
  \BibitemOpen
  \bibfield  {author} {\bibinfo {author} {\bibfnamefont {Y.}~\bibnamefont
  {Cao}}, \bibinfo {author} {\bibfnamefont {V.}~\bibnamefont {Fatemi}},
  \bibinfo {author} {\bibfnamefont {S.}~\bibnamefont {Fang}}, \bibinfo {author}
  {\bibfnamefont {K.}~\bibnamefont {Watanabe}}, \bibinfo {author}
  {\bibfnamefont {T.}~\bibnamefont {Taniguchi}}, \bibinfo {author}
  {\bibfnamefont {E.}~\bibnamefont {Kaxiras}},\ and\ \bibinfo {author}
  {\bibfnamefont {P.}~\bibnamefont {Jarillo-Herrero}},\ }\bibfield  {title}
  {\bibinfo {title} {Unconventional superconductivity in magic-angle graphene
  superlattices},\ }\href {https://doi.org/10.1038/nature26160} {\bibfield
  {journal} {\bibinfo  {journal} {Nature}\ }\textbf {\bibinfo {volume} {556}},\
  \bibinfo {pages} {43} (\bibinfo {year} {2018})}\BibitemShut {NoStop}%
\bibitem [{\citenamefont {Sajadi}\ \emph {et~al.}(2018)\citenamefont {Sajadi},
  \citenamefont {Palomaki}, \citenamefont {Fei}, \citenamefont {Zhao},
  \citenamefont {Bement}, \citenamefont {Olsen}, \citenamefont {Luescher},
  \citenamefont {Xu}, \citenamefont {Folk},\ and\ \citenamefont
  {Cobden}}]{sajadi_gate-induced_2018}%
  \BibitemOpen
  \bibfield  {author} {\bibinfo {author} {\bibfnamefont {E.}~\bibnamefont
  {Sajadi}}, \bibinfo {author} {\bibfnamefont {T.}~\bibnamefont {Palomaki}},
  \bibinfo {author} {\bibfnamefont {Z.}~\bibnamefont {Fei}}, \bibinfo {author}
  {\bibfnamefont {W.}~\bibnamefont {Zhao}}, \bibinfo {author} {\bibfnamefont
  {P.}~\bibnamefont {Bement}}, \bibinfo {author} {\bibfnamefont
  {C.}~\bibnamefont {Olsen}}, \bibinfo {author} {\bibfnamefont
  {S.}~\bibnamefont {Luescher}}, \bibinfo {author} {\bibfnamefont
  {X.}~\bibnamefont {Xu}}, \bibinfo {author} {\bibfnamefont {J.~A.}\
  \bibnamefont {Folk}},\ and\ \bibinfo {author} {\bibfnamefont {D.~H.}\
  \bibnamefont {Cobden}},\ }\bibfield  {title} {\bibinfo {title} {Gate-induced
  superconductivity in a monolayer topological insulator},\ }\href
  {https://doi.org/10.1126/science.aar4426} {\bibfield  {journal} {\bibinfo
  {journal} {Science}\ }\textbf {\bibinfo {volume} {362}},\ \bibinfo {pages}
  {922} (\bibinfo {year} {2018})}\BibitemShut {NoStop}%
\bibitem [{\citenamefont {Fatemi}\ \emph {et~al.}(2018)\citenamefont {Fatemi},
  \citenamefont {Wu}, \citenamefont {Cao}, \citenamefont {Bretheau},
  \citenamefont {Gibson}, \citenamefont {Watanabe}, \citenamefont {Taniguchi},
  \citenamefont {Cava},\ and\ \citenamefont
  {Jarillo-Herrero}}]{fatemi_electrically_2018}%
  \BibitemOpen
  \bibfield  {author} {\bibinfo {author} {\bibfnamefont {V.}~\bibnamefont
  {Fatemi}}, \bibinfo {author} {\bibfnamefont {S.}~\bibnamefont {Wu}}, \bibinfo
  {author} {\bibfnamefont {Y.}~\bibnamefont {Cao}}, \bibinfo {author}
  {\bibfnamefont {L.}~\bibnamefont {Bretheau}}, \bibinfo {author}
  {\bibfnamefont {Q.~D.}\ \bibnamefont {Gibson}}, \bibinfo {author}
  {\bibfnamefont {K.}~\bibnamefont {Watanabe}}, \bibinfo {author}
  {\bibfnamefont {T.}~\bibnamefont {Taniguchi}}, \bibinfo {author}
  {\bibfnamefont {R.~J.}\ \bibnamefont {Cava}},\ and\ \bibinfo {author}
  {\bibfnamefont {P.}~\bibnamefont {Jarillo-Herrero}},\ }\bibfield  {title}
  {\bibinfo {title} {Electrically tunable low-density superconductivity in a
  monolayer topological insulator},\ }\href
  {https://doi.org/10.1126/science.aar4642} {\bibfield  {journal} {\bibinfo
  {journal} {Science}\ }\textbf {\bibinfo {volume} {362}},\ \bibinfo {pages}
  {926} (\bibinfo {year} {2018})}\BibitemShut {NoStop}%
\bibitem [{\citenamefont {Mathur}\ \emph {et~al.}(2010)\citenamefont {Mathur},
  \citenamefont {Raychowdhury},\ and\ \citenamefont {Anishetty}}]{mathur2010n}%
  \BibitemOpen
  \bibfield  {author} {\bibinfo {author} {\bibfnamefont {M.}~\bibnamefont
  {Mathur}}, \bibinfo {author} {\bibfnamefont {I.}~\bibnamefont
  {Raychowdhury}},\ and\ \bibinfo {author} {\bibfnamefont {R.}~\bibnamefont
  {Anishetty}},\ }\bibfield  {title} {\bibinfo {title} {Su (n) irreducible
  schwinger bosons},\ }\href@noop {} {\bibfield  {journal} {\bibinfo  {journal}
  {Journal of mathematical physics}\ }\textbf {\bibinfo {volume} {51}},\
  \bibinfo {pages} {093504} (\bibinfo {year} {2010})}\BibitemShut {NoStop}%
\bibitem [{\citenamefont {Cram{\'e}r}(1999)}]{cramer1999mathematical}%
  \BibitemOpen
  \bibfield  {author} {\bibinfo {author} {\bibfnamefont {H.}~\bibnamefont
  {Cram{\'e}r}},\ }\href@noop {} {\emph {\bibinfo {title} {Mathematical methods
  of statistics}}},\ Vol.~\bibinfo {volume} {26}\ (\bibinfo  {publisher}
  {Princeton university press},\ \bibinfo {year} {1999})\BibitemShut {NoStop}%
\bibitem [{\citenamefont {Rao}\ \emph {et~al.}(1992)\citenamefont {Rao} \emph
  {et~al.}}]{rao1992information}%
  \BibitemOpen
  \bibfield  {author} {\bibinfo {author} {\bibfnamefont {C.~R.}\ \bibnamefont
  {Rao}} \emph {et~al.},\ }\bibfield  {title} {\bibinfo {title} {Information
  and the accuracy attainable in the estimation of statistical parameters},\
  }\href@noop {} {\bibfield  {journal} {\bibinfo  {journal} {Breakthroughs in
  statistics}\ ,\ \bibinfo {pages} {235}} (\bibinfo {year} {1992})}\BibitemShut
  {NoStop}%
\bibitem [{\citenamefont {Sidhu}\ and\ \citenamefont
  {Kok}(2020)}]{sidhu2020geometric}%
  \BibitemOpen
  \bibfield  {author} {\bibinfo {author} {\bibfnamefont {J.~S.}\ \bibnamefont
  {Sidhu}}\ and\ \bibinfo {author} {\bibfnamefont {P.}~\bibnamefont {Kok}},\
  }\bibfield  {title} {\bibinfo {title} {Geometric perspective on quantum
  parameter estimation},\ }\href@noop {} {\bibfield  {journal} {\bibinfo
  {journal} {AVS Quantum Science}\ }\textbf {\bibinfo {volume} {2}},\ \bibinfo
  {pages} {014701} (\bibinfo {year} {2020})}\BibitemShut {NoStop}%
\bibitem [{\citenamefont {Yu}\ \emph {et~al.}(2022)\citenamefont {Yu},
  \citenamefont {Liu}, \citenamefont {Yang}, \citenamefont {Gong},
  \citenamefont {Cao}, \citenamefont {Zhang}, \citenamefont {Liu},
  \citenamefont {Heyl}, \citenamefont {Ozawa}, \citenamefont {Goldman} \emph
  {et~al.}}]{yu2022quantum}%
  \BibitemOpen
  \bibfield  {author} {\bibinfo {author} {\bibfnamefont {M.}~\bibnamefont
  {Yu}}, \bibinfo {author} {\bibfnamefont {Y.}~\bibnamefont {Liu}}, \bibinfo
  {author} {\bibfnamefont {P.}~\bibnamefont {Yang}}, \bibinfo {author}
  {\bibfnamefont {M.}~\bibnamefont {Gong}}, \bibinfo {author} {\bibfnamefont
  {Q.}~\bibnamefont {Cao}}, \bibinfo {author} {\bibfnamefont {S.}~\bibnamefont
  {Zhang}}, \bibinfo {author} {\bibfnamefont {H.}~\bibnamefont {Liu}}, \bibinfo
  {author} {\bibfnamefont {M.}~\bibnamefont {Heyl}}, \bibinfo {author}
  {\bibfnamefont {T.}~\bibnamefont {Ozawa}}, \bibinfo {author} {\bibfnamefont
  {N.}~\bibnamefont {Goldman}}, \emph {et~al.},\ }\bibfield  {title} {\bibinfo
  {title} {Quantum fisher information measurement and verification of the
  quantum cram{\'e}r--rao bound in a solid-state qubit},\ }\href@noop {}
  {\bibfield  {journal} {\bibinfo  {journal} {npj Quantum Information}\
  }\textbf {\bibinfo {volume} {8}},\ \bibinfo {pages} {56} (\bibinfo {year}
  {2022})}\BibitemShut {NoStop}%
\bibitem [{\citenamefont {Takahashi}\ \emph {et~al.}(2013)\citenamefont
  {Takahashi}, \citenamefont {Bartlett},\ and\ \citenamefont
  {Doherty}}]{takahashi2013tomography}%
  \BibitemOpen
  \bibfield  {author} {\bibinfo {author} {\bibfnamefont {M.}~\bibnamefont
  {Takahashi}}, \bibinfo {author} {\bibfnamefont {S.~D.}\ \bibnamefont
  {Bartlett}},\ and\ \bibinfo {author} {\bibfnamefont {A.~C.}\ \bibnamefont
  {Doherty}},\ }\bibfield  {title} {\bibinfo {title} {Tomography of a spin
  qubit in a double quantum dot},\ }\href@noop {} {\bibfield  {journal}
  {\bibinfo  {journal} {Physical Review A}\ }\textbf {\bibinfo {volume} {88}},\
  \bibinfo {pages} {022120} (\bibinfo {year} {2013})}\BibitemShut {NoStop}%
\bibitem [{\citenamefont {Chuang}\ and\ \citenamefont
  {Nielsen}(1997)}]{chuang1997prescription}%
  \BibitemOpen
  \bibfield  {author} {\bibinfo {author} {\bibfnamefont {I.~L.}\ \bibnamefont
  {Chuang}}\ and\ \bibinfo {author} {\bibfnamefont {M.~A.}\ \bibnamefont
  {Nielsen}},\ }\bibfield  {title} {\bibinfo {title} {Prescription for
  experimental determination of the dynamics of a quantum black box},\
  }\href@noop {} {\bibfield  {journal} {\bibinfo  {journal} {Journal of Modern
  Optics}\ }\textbf {\bibinfo {volume} {44}},\ \bibinfo {pages} {2455}
  (\bibinfo {year} {1997})}\BibitemShut {NoStop}%
\bibitem [{\citenamefont {Sturges}\ \emph {et~al.}(2021)\citenamefont
  {Sturges}, \citenamefont {McDermott}, \citenamefont {Buraczewski},
  \citenamefont {Clements}, \citenamefont {Renema}, \citenamefont {Nam},
  \citenamefont {Gerrits}, \citenamefont {Lita}, \citenamefont {Kolthammer},
  \citenamefont {Eckstein}, \citenamefont {Walmsley},\ and\ \citenamefont
  {Stobińska}}]{sturges_quantum_2021}%
  \BibitemOpen
  \bibfield  {author} {\bibinfo {author} {\bibfnamefont {T.~J.}\ \bibnamefont
  {Sturges}}, \bibinfo {author} {\bibfnamefont {T.}~\bibnamefont {McDermott}},
  \bibinfo {author} {\bibfnamefont {A.}~\bibnamefont {Buraczewski}}, \bibinfo
  {author} {\bibfnamefont {W.~R.}\ \bibnamefont {Clements}}, \bibinfo {author}
  {\bibfnamefont {J.~J.}\ \bibnamefont {Renema}}, \bibinfo {author}
  {\bibfnamefont {S.~W.}\ \bibnamefont {Nam}}, \bibinfo {author} {\bibfnamefont
  {T.}~\bibnamefont {Gerrits}}, \bibinfo {author} {\bibfnamefont
  {A.}~\bibnamefont {Lita}}, \bibinfo {author} {\bibfnamefont {W.~S.}\
  \bibnamefont {Kolthammer}}, \bibinfo {author} {\bibfnamefont
  {A.}~\bibnamefont {Eckstein}}, \bibinfo {author} {\bibfnamefont {I.~A.}\
  \bibnamefont {Walmsley}},\ and\ \bibinfo {author} {\bibfnamefont
  {M.}~\bibnamefont {Stobińska}},\ }\bibfield  {title} {\bibinfo {title}
  {Quantum simulations with multiphoton {Fock} states},\ }\href
  {https://doi.org/10.1038/s41534-021-00427-w} {\bibfield  {journal} {\bibinfo
  {journal} {npj Quantum Information}\ }\textbf {\bibinfo {volume} {7}},\
  \bibinfo {pages} {91} (\bibinfo {year} {2021})}\BibitemShut {NoStop}%
\bibitem [{\citenamefont {Brown}\ \emph {et~al.}(2003)\citenamefont {Brown},
  \citenamefont {Dani}, \citenamefont {Stamper-Kurn},\ and\ \citenamefont
  {Whaley}}]{brown_deterministic_2003}%
  \BibitemOpen
  \bibfield  {author} {\bibinfo {author} {\bibfnamefont {K.~R.}\ \bibnamefont
  {Brown}}, \bibinfo {author} {\bibfnamefont {K.~M.}\ \bibnamefont {Dani}},
  \bibinfo {author} {\bibfnamefont {D.~M.}\ \bibnamefont {Stamper-Kurn}},\ and\
  \bibinfo {author} {\bibfnamefont {K.~B.}\ \bibnamefont {Whaley}},\ }\bibfield
   {title} {\bibinfo {title} {Deterministic optical {Fock}-state generation},\
  }\href {https://doi.org/10.1103/PhysRevA.67.043818} {\bibfield  {journal}
  {\bibinfo  {journal} {Physical Review A}\ }\textbf {\bibinfo {volume} {67}},\
  \bibinfo {pages} {043818} (\bibinfo {year} {2003})}\BibitemShut {NoStop}%
\bibitem [{\citenamefont {Velez}\ \emph {et~al.}(2019)\citenamefont {Velez},
  \citenamefont {Seibold}, \citenamefont {Kipfer}, \citenamefont {Anderson},
  \citenamefont {Sudhir},\ and\ \citenamefont
  {Galland}}]{velez_preparation_2019}%
  \BibitemOpen
  \bibfield  {author} {\bibinfo {author} {\bibfnamefont {S.~T.}\ \bibnamefont
  {Velez}}, \bibinfo {author} {\bibfnamefont {K.}~\bibnamefont {Seibold}},
  \bibinfo {author} {\bibfnamefont {N.}~\bibnamefont {Kipfer}}, \bibinfo
  {author} {\bibfnamefont {M.~D.}\ \bibnamefont {Anderson}}, \bibinfo {author}
  {\bibfnamefont {V.}~\bibnamefont {Sudhir}},\ and\ \bibinfo {author}
  {\bibfnamefont {C.}~\bibnamefont {Galland}},\ }\bibfield  {title} {\bibinfo
  {title} {Preparation and {Decay} of a {Single} {Quantum} of {Vibration} at
  {Ambient} {Conditions}},\ }\href {https://doi.org/10.1103/PhysRevX.9.041007}
  {\bibfield  {journal} {\bibinfo  {journal} {Physical Review X}\ }\textbf
  {\bibinfo {volume} {9}},\ \bibinfo {pages} {041007} (\bibinfo {year}
  {2019})}\BibitemShut {NoStop}%
\bibitem [{\citenamefont {Henderson}\ \emph {et~al.}(2008)\citenamefont
  {Henderson}, \citenamefont {Ramsey}, \citenamefont {Quddusi},\ and\
  \citenamefont {del Barco}}]{henderson_high-frequency_2008}%
  \BibitemOpen
  \bibfield  {author} {\bibinfo {author} {\bibfnamefont {J.~J.}\ \bibnamefont
  {Henderson}}, \bibinfo {author} {\bibfnamefont {C.~M.}\ \bibnamefont
  {Ramsey}}, \bibinfo {author} {\bibfnamefont {H.~M.}\ \bibnamefont
  {Quddusi}},\ and\ \bibinfo {author} {\bibfnamefont {E.}~\bibnamefont {del
  Barco}},\ }\bibfield  {title} {\bibinfo {title} {High-frequency microstrip
  cross resonators for circular polarization electron paramagnetic resonance
  spectroscopy},\ }\href {https://doi.org/10.1063/1.2957621} {\bibfield
  {journal} {\bibinfo  {journal} {Review of Scientific Instruments}\ }\textbf
  {\bibinfo {volume} {79}},\ \bibinfo {pages} {074704} (\bibinfo {year}
  {2008})}\BibitemShut {NoStop}%
\bibitem [{\citenamefont {Alegre}\ \emph {et~al.}(2007)\citenamefont {Alegre},
  \citenamefont {Santori}, \citenamefont {Medeiros-Ribeiro},\ and\
  \citenamefont {Beausoleil}}]{alegre_polarization-selective_2007}%
  \BibitemOpen
  \bibfield  {author} {\bibinfo {author} {\bibfnamefont {T.~P.~M.}\
  \bibnamefont {Alegre}}, \bibinfo {author} {\bibfnamefont {C.}~\bibnamefont
  {Santori}}, \bibinfo {author} {\bibfnamefont {G.}~\bibnamefont
  {Medeiros-Ribeiro}},\ and\ \bibinfo {author} {\bibfnamefont {R.~G.}\
  \bibnamefont {Beausoleil}},\ }\bibfield  {title} {\bibinfo {title}
  {Polarization-selective excitation of nitrogen vacancy centers in diamond},\
  }\href {https://doi.org/10.1103/PhysRevB.76.165205} {\bibfield  {journal}
  {\bibinfo  {journal} {Physical Review B}\ }\textbf {\bibinfo {volume} {76}},\
  \bibinfo {pages} {165205} (\bibinfo {year} {2007})}\BibitemShut {NoStop}%
\bibitem [{\citenamefont {Romanenko}\ \emph {et~al.}(2020)\citenamefont
  {Romanenko}, \citenamefont {Pilipenko}, \citenamefont {Zorzetti},
  \citenamefont {Frolov}, \citenamefont {Awida}, \citenamefont {Belomestnykh},
  \citenamefont {Posen},\ and\ \citenamefont
  {Grassellino}}]{romanenko_three-dimensional_2020}%
  \BibitemOpen
  \bibfield  {author} {\bibinfo {author} {\bibfnamefont {A.}~\bibnamefont
  {Romanenko}}, \bibinfo {author} {\bibfnamefont {R.}~\bibnamefont
  {Pilipenko}}, \bibinfo {author} {\bibfnamefont {S.}~\bibnamefont {Zorzetti}},
  \bibinfo {author} {\bibfnamefont {D.}~\bibnamefont {Frolov}}, \bibinfo
  {author} {\bibfnamefont {M.}~\bibnamefont {Awida}}, \bibinfo {author}
  {\bibfnamefont {S.}~\bibnamefont {Belomestnykh}}, \bibinfo {author}
  {\bibfnamefont {S.}~\bibnamefont {Posen}},\ and\ \bibinfo {author}
  {\bibfnamefont {A.}~\bibnamefont {Grassellino}},\ }\bibfield  {title}
  {\bibinfo {title} {Three-{Dimensional} {Superconducting} {Resonators} at {T}
  {\textless} 20 {mK} with {Photon} {Lifetimes} up to $\tau$ = 2 s},\ }\href
  {https://doi.org/10.1103/PhysRevApplied.13.034032} {\bibfield  {journal}
  {\bibinfo  {journal} {Physical Review Applied}\ }\textbf {\bibinfo {volume}
  {13}},\ \bibinfo {pages} {034032} (\bibinfo {year} {2020})}\BibitemShut
  {NoStop}%
\bibitem [{\citenamefont {Hofheinz}\ \emph {et~al.}(2008)\citenamefont
  {Hofheinz}, \citenamefont {Weig}, \citenamefont {Ansmann}, \citenamefont
  {Bialczak}, \citenamefont {Lucero}, \citenamefont {Neeley}, \citenamefont
  {O’Connell}, \citenamefont {Wang}, \citenamefont {Martinis},\ and\
  \citenamefont {Cleland}}]{hofheinz_generation_2008}%
  \BibitemOpen
  \bibfield  {author} {\bibinfo {author} {\bibfnamefont {M.}~\bibnamefont
  {Hofheinz}}, \bibinfo {author} {\bibfnamefont {E.~M.}\ \bibnamefont {Weig}},
  \bibinfo {author} {\bibfnamefont {M.}~\bibnamefont {Ansmann}}, \bibinfo
  {author} {\bibfnamefont {R.~C.}\ \bibnamefont {Bialczak}}, \bibinfo {author}
  {\bibfnamefont {E.}~\bibnamefont {Lucero}}, \bibinfo {author} {\bibfnamefont
  {M.}~\bibnamefont {Neeley}}, \bibinfo {author} {\bibfnamefont {A.~D.}\
  \bibnamefont {O’Connell}}, \bibinfo {author} {\bibfnamefont
  {H.}~\bibnamefont {Wang}}, \bibinfo {author} {\bibfnamefont {J.~M.}\
  \bibnamefont {Martinis}},\ and\ \bibinfo {author} {\bibfnamefont {A.~N.}\
  \bibnamefont {Cleland}},\ }\bibfield  {title} {\bibinfo {title} {Generation
  of {Fock} states in a superconducting quantum circuit},\ }\href
  {https://doi.org/10.1038/nature07136} {\bibfield  {journal} {\bibinfo
  {journal} {Nature}\ }\textbf {\bibinfo {volume} {454}},\ \bibinfo {pages}
  {310} (\bibinfo {year} {2008})}\BibitemShut {NoStop}%
\bibitem [{\citenamefont {Wang}\ \emph {et~al.}(2020)\citenamefont {Wang},
  \citenamefont {Curtis}, \citenamefont {Lester}, \citenamefont {Zhang},
  \citenamefont {Gao}, \citenamefont {Freeze}, \citenamefont {Batista},
  \citenamefont {Vaccaro}, \citenamefont {Chuang}, \citenamefont {Frunzio},
  \citenamefont {Jiang}, \citenamefont {Girvin},\ and\ \citenamefont
  {Schoelkopf}}]{wang_efficient_2020}%
  \BibitemOpen
  \bibfield  {author} {\bibinfo {author} {\bibfnamefont {C.~S.}\ \bibnamefont
  {Wang}}, \bibinfo {author} {\bibfnamefont {J.~C.}\ \bibnamefont {Curtis}},
  \bibinfo {author} {\bibfnamefont {B.~J.}\ \bibnamefont {Lester}}, \bibinfo
  {author} {\bibfnamefont {Y.}~\bibnamefont {Zhang}}, \bibinfo {author}
  {\bibfnamefont {Y.~Y.}\ \bibnamefont {Gao}}, \bibinfo {author} {\bibfnamefont
  {J.}~\bibnamefont {Freeze}}, \bibinfo {author} {\bibfnamefont {V.~S.}\
  \bibnamefont {Batista}}, \bibinfo {author} {\bibfnamefont {P.~H.}\
  \bibnamefont {Vaccaro}}, \bibinfo {author} {\bibfnamefont {I.~L.}\
  \bibnamefont {Chuang}}, \bibinfo {author} {\bibfnamefont {L.}~\bibnamefont
  {Frunzio}}, \bibinfo {author} {\bibfnamefont {L.}~\bibnamefont {Jiang}},
  \bibinfo {author} {\bibfnamefont {S.}~\bibnamefont {Girvin}},\ and\ \bibinfo
  {author} {\bibfnamefont {R.~J.}\ \bibnamefont {Schoelkopf}},\ }\bibfield
  {title} {\bibinfo {title} {Efficient {Multiphoton} {Sampling} of {Molecular}
  {Vibronic} {Spectra} on a {Superconducting} {Bosonic} {Processor}},\ }\href
  {https://doi.org/10.1103/PhysRevX.10.021060} {\bibfield  {journal} {\bibinfo
  {journal} {Physical Review X}\ }\textbf {\bibinfo {volume} {10}},\ \bibinfo
  {pages} {021060} (\bibinfo {year} {2020})}\BibitemShut {NoStop}%
\bibitem [{\citenamefont {Gao}\ \emph {et~al.}(2018)\citenamefont {Gao},
  \citenamefont {Lester}, \citenamefont {Zhang}, \citenamefont {Wang},
  \citenamefont {Rosenblum}, \citenamefont {Frunzio}, \citenamefont {Jiang},
  \citenamefont {Girvin},\ and\ \citenamefont
  {Schoelkopf}}]{gao_programmable_2018}%
  \BibitemOpen
  \bibfield  {author} {\bibinfo {author} {\bibfnamefont {Y.~Y.}\ \bibnamefont
  {Gao}}, \bibinfo {author} {\bibfnamefont {B.~J.}\ \bibnamefont {Lester}},
  \bibinfo {author} {\bibfnamefont {Y.}~\bibnamefont {Zhang}}, \bibinfo
  {author} {\bibfnamefont {C.}~\bibnamefont {Wang}}, \bibinfo {author}
  {\bibfnamefont {S.}~\bibnamefont {Rosenblum}}, \bibinfo {author}
  {\bibfnamefont {L.}~\bibnamefont {Frunzio}}, \bibinfo {author} {\bibfnamefont
  {L.}~\bibnamefont {Jiang}}, \bibinfo {author} {\bibfnamefont
  {S.}~\bibnamefont {Girvin}},\ and\ \bibinfo {author} {\bibfnamefont {R.~J.}\
  \bibnamefont {Schoelkopf}},\ }\bibfield  {title} {\bibinfo {title}
  {Programmable {Interference} between {Two} {Microwave} {Quantum}
  {Memories}},\ }\href {https://doi.org/10.1103/PhysRevX.8.021073} {\bibfield
  {journal} {\bibinfo  {journal} {Physical Review X}\ }\textbf {\bibinfo
  {volume} {8}},\ \bibinfo {pages} {021073} (\bibinfo {year}
  {2018})}\BibitemShut {NoStop}%
\bibitem [{\citenamefont {Matthews}\ \emph {et~al.}(2016)\citenamefont
  {Matthews}, \citenamefont {Zhou}, \citenamefont {Cable}, \citenamefont
  {Shadbolt}, \citenamefont {Saunders}, \citenamefont {Durkin}, \citenamefont
  {Pryde},\ and\ \citenamefont {O’Brien}}]{matthews2016towards}%
  \BibitemOpen
  \bibfield  {author} {\bibinfo {author} {\bibfnamefont {J.~C.}\ \bibnamefont
  {Matthews}}, \bibinfo {author} {\bibfnamefont {X.-Q.}\ \bibnamefont {Zhou}},
  \bibinfo {author} {\bibfnamefont {H.}~\bibnamefont {Cable}}, \bibinfo
  {author} {\bibfnamefont {P.~J.}\ \bibnamefont {Shadbolt}}, \bibinfo {author}
  {\bibfnamefont {D.~J.}\ \bibnamefont {Saunders}}, \bibinfo {author}
  {\bibfnamefont {G.~A.}\ \bibnamefont {Durkin}}, \bibinfo {author}
  {\bibfnamefont {G.~J.}\ \bibnamefont {Pryde}},\ and\ \bibinfo {author}
  {\bibfnamefont {J.~L.}\ \bibnamefont {O’Brien}},\ }\bibfield  {title}
  {\bibinfo {title} {Towards practical quantum metrology with photon
  counting},\ }\href@noop {} {\bibfield  {journal} {\bibinfo  {journal} {npj
  Quantum Information}\ }\textbf {\bibinfo {volume} {2}},\ \bibinfo {pages} {1}
  (\bibinfo {year} {2016})}\BibitemShut {NoStop}%
\bibitem [{\citenamefont {Xia}\ \emph {et~al.}(2015)\citenamefont {Xia},
  \citenamefont {Zhao}, \citenamefont {Twamley},\ and\ \citenamefont {{EQuS
  Collaboration}}}]{xia_detection_2015}%
  \BibitemOpen
  \bibfield  {author} {\bibinfo {author} {\bibfnamefont {K.}~\bibnamefont
  {Xia}}, \bibinfo {author} {\bibfnamefont {N.}~\bibnamefont {Zhao}}, \bibinfo
  {author} {\bibfnamefont {J.}~\bibnamefont {Twamley}},\ and\ \bibinfo {author}
  {\bibnamefont {{EQuS Collaboration}}},\ }\bibfield  {title} {\bibinfo {title}
  {Detection of a weak magnetic field via cavity-enhanced {Faraday} rotation},\
  }\href {https://doi.org/10.1103/PhysRevA.92.043409} {\bibfield  {journal}
  {\bibinfo  {journal} {Physical Review A}\ }\textbf {\bibinfo {volume} {92}},\
  \bibinfo {pages} {043409} (\bibinfo {year} {2015})}\BibitemShut {NoStop}%
\bibitem [{\citenamefont {Lutchyn}\ \emph {et~al.}(2009)\citenamefont
  {Lutchyn}, \citenamefont {Nagornykh},\ and\ \citenamefont
  {Yakovenko}}]{lutchyn_frequency_2009}%
  \BibitemOpen
  \bibfield  {author} {\bibinfo {author} {\bibfnamefont {R.~M.}\ \bibnamefont
  {Lutchyn}}, \bibinfo {author} {\bibfnamefont {P.}~\bibnamefont {Nagornykh}},\
  and\ \bibinfo {author} {\bibfnamefont {V.~M.}\ \bibnamefont {Yakovenko}},\
  }\bibfield  {title} {\bibinfo {title} {Frequency and temperature dependence
  of the anomalous ac {Hall} conductivity in a chiral p x + i p y
  superconductor with impurities},\ }\href
  {https://doi.org/10.1103/PhysRevB.80.104508} {\bibfield  {journal} {\bibinfo
  {journal} {Physical Review B}\ }\textbf {\bibinfo {volume} {80}},\ \bibinfo
  {pages} {104508} (\bibinfo {year} {2009})}\BibitemShut {NoStop}%
\bibitem [{\citenamefont {Hayashi}\ \emph {et~al.}(2021)\citenamefont
  {Hayashi}, \citenamefont {Okamura}, \citenamefont {Kanazawa}, \citenamefont
  {Yu}, \citenamefont {Koretsune}, \citenamefont {Arita}, \citenamefont
  {Tsukazaki}, \citenamefont {Ichikawa}, \citenamefont {Kawasaki},
  \citenamefont {Tokura},\ and\ \citenamefont
  {Takahashi}}]{hayashi_magneto-optical_2021}%
  \BibitemOpen
  \bibfield  {author} {\bibinfo {author} {\bibfnamefont {Y.}~\bibnamefont
  {Hayashi}}, \bibinfo {author} {\bibfnamefont {Y.}~\bibnamefont {Okamura}},
  \bibinfo {author} {\bibfnamefont {N.}~\bibnamefont {Kanazawa}}, \bibinfo
  {author} {\bibfnamefont {T.}~\bibnamefont {Yu}}, \bibinfo {author}
  {\bibfnamefont {T.}~\bibnamefont {Koretsune}}, \bibinfo {author}
  {\bibfnamefont {R.}~\bibnamefont {Arita}}, \bibinfo {author} {\bibfnamefont
  {A.}~\bibnamefont {Tsukazaki}}, \bibinfo {author} {\bibfnamefont
  {M.}~\bibnamefont {Ichikawa}}, \bibinfo {author} {\bibfnamefont
  {M.}~\bibnamefont {Kawasaki}}, \bibinfo {author} {\bibfnamefont
  {Y.}~\bibnamefont {Tokura}},\ and\ \bibinfo {author} {\bibfnamefont
  {Y.}~\bibnamefont {Takahashi}},\ }\bibfield  {title} {\bibinfo {title}
  {Magneto-optical spectroscopy on {Weyl} nodes for anomalous and topological
  {Hall} effects in chiral {MnGe}},\ }\href
  {https://doi.org/10.1038/s41467-021-25276-1} {\bibfield  {journal} {\bibinfo
  {journal} {Nature Communications}\ }\textbf {\bibinfo {volume} {12}},\
  \bibinfo {pages} {5974} (\bibinfo {year} {2021})}\BibitemShut {NoStop}%
\end{thebibliography}
%

\end{document}